\documentclass{article}

\usepackage{arxiv}

\usepackage[utf8]{inputenc} 
\usepackage[T1]{fontenc}    
\usepackage{hyperref}       
\usepackage{url}            
\usepackage{booktabs}       
\usepackage{amsfonts}       
\usepackage{nicefrac}       
\usepackage{microtype}      
\usepackage{lipsum}		
\usepackage{graphicx}
\usepackage{doi}
\usepackage{amssymb}
\usepackage{amssymb,amsmath}
\usepackage{mathtools}
\usepackage{eqnarray}
\usepackage{IEEEtrantools}
\usepackage{multirow}
\usepackage{placeins}
\usepackage{svg}
\usepackage{booktabs} 
\usepackage{float}
\usepackage{graphicx}
\usepackage{subfig}
\usepackage{bm}
\usepackage{enumerate}
\usepackage{enumitem}
\usepackage{soul}
\usepackage{setspace}
\usepackage{textcomp}
\usepackage{graphicx} 
\usepackage{subfig}

\usepackage{todonotes}
\usepackage{txfonts}
\usepackage{dsfont}
\newlist{steps}{enumerate}{1}
\setlist[steps, 1]{label = \textbf{Step \arabic*:},leftmargin=1.3cm}
\usepackage{cleveref}
\usepackage{caption}
\usepackage{lipsum}
\newcommand{\Hf}{ \mathcal{H}} 

\newcommand{\fullG}{\bm{ \Upsilon}}

\newcommand{\ys}{\boldsymbol{y}}
\newcommand{\obs}{\varphi}
\newcommand{\Obs}{\Phi}

\newcommand{\frf}{{\small{\textsf{FRF}}}}

\include{pythonlisting}

\usepackage{algorithmic}
\usepackage{algorithm}
\usepackage{bbm}
\renewcommand{\algorithmicrequire}{\textbf{Input:}}
\renewcommand{\algorithmicensure}{\textbf{Output:}}

\usepackage[switch, modulo]{lineno}

\title{A wavelet-based dynamic mode decomposition for modeling mechanical systems from partial observations}

\author{{Manu Krishnan \thanks{Address all correspondence to this author}} \\
	Department of Aerospace and Ocean Engineering \\
	Virginia Polytechnic Institute and State University\\
 Blacksburg, Virginia - 24060 \\
	\texttt{manukris@vt.edu} \\
	\And
{Serkan Gugercin} \\
	Department of Mathematics and Division of Computational Modeling \& Data Analytics\\
	Virginia Polytechnic Institute and State University\\
 Blacksburg, Virginia - 24060 \\
 \texttt{gugercin@vt.edu} \\
	\And
{Pablo A. Tarazaga \thanks{Affliated with Department of Mechanical Engineering, Virginia Polytechnic Institute and State University}}\\
	Department of Mechanical Engineering \\
		Texas A \& M\\
	College Station, Texas - 77843\\
	\texttt{ptarazaga@tamu.edu} 
	}

\renewcommand{\shorttitle}{Wavelet-based DMD}

\hypersetup{
pdftitle={A template for the arxiv style},
pdfsubject={q-bio.NC, q-bio.QM},
pdfauthor={David S.~Hippocampus, Elias D.~Striatum},
pdfkeywords={First keyword, Second keyword, More},
}

\begin{document}
\maketitle

\begin{abstract}
\normalsize
Dynamic mode decomposition (DMD) has emerged as a popular data-driven modeling approach to identifying  spatio-temporal coherent structures in dynamical systems, owing to its strong relation with the Koopman operator. For dynamical systems with external forcing, the identified model should not only be suitable for a specific forcing function but should generally approximate the input-output behavior of
the underlying dynamics.  A novel methodology for modeling those classes of dynamical systems is proposed in the present work, using wavelets in conjunction with the input-output dynamic mode decomposition (ioDMD). The wavelet-based dynamic mode decomposition (WDMD) builds on the ioDMD framework without the restrictive assumption of full state measurements. Our non-intrusive approach constructs numerical models directly from trajectories of the full model's inputs and outputs, without requiring the full-model operators.
These trajectories are generated by running a simulation of the full model or observing the original dynamical systems' response to inputs in an experimental framework. Hence, the present methodology is applicable for dynamical systems whose internal state vector measurements are not available. Instead, data from only a few output locations are only accessible, as often the case in practice. The present methodology's applicability is explained by modeling the input-output response of an Euler-Bernoulli finite element beam model. The WDMD provides a linear state-space representation of the dynamical system using the response measurements and the corresponding input forcing functions. The developed state-space model can then be used to simulate the beam's response towards different types of forcing functions. The method is further validated on a real (experimental) data set using  modal analysis on a simple free-free beam, demonstrating the efficacy of the proposed
methodology as an appropriate candidate for modeling practical dynamical systems despite having no access to internal state measurements and treating the full model as a black-box. 

\end{abstract}

\section{INTRODUCTION}
Over the last two decades, data-driven modeling has garnered interest in several research areas, particularly when the involved dynamics are complex and models based on first principles present challenges of varying degrees 
 {see, e.g., \mbox{\cite{schmid2010dynamic, kutz2016dynamic,  dmdbook, qian2020lift, brunton2019data, ghattas2021learning,loewnerchapter,AntBG20,modredvol1}} and the references therein}. Moreover, the advances in data processing and sensor capabilities made it much easier to map a system's response to a variety of inputs. 
In  structural dynamics, structures vary in different levels of complexity, and physics-based models may not always be feasible \cite{souza2020bayesian, bertha2016multivariate, el2019mlmm, formenti2002parameter, guillaume2003poly, 
kim2019response, albakri2020estimating, malladi2021estimating, krishnan2022leveraging}.
To gain insight into the aforementioned class of dynamical systems and predict their behaviors in different operational or loading conditions, it is beneficial to create models using the measured input-output response data.

Dynamic mode decomposition (DMD) has become a popular tool in data-driven modeling, owing to its ability to decompose the high dimensional data into its coherent spatio-temporal structures \cite{schmid2010dynamic, kutz2016dynamic}. DMD has its roots in the Koopman theory \cite{koopman1931hamiltonian}, whose work was later revived by seminal works of Mezi\'{c} et al. \cite{ref1, rowley2009spectral}. Koopman theory in essence, associates a nonlinear dynamical system with an infinite-dimensional linear system,  allowing tools for linear systems  theory to be employed. Different variants of DMD have been proposed to improve the pre-processing or post-processing capabilities of the standard DMD. Optimized DMD \cite{chen2012variants} and sparsity promoting DMD \cite{jovanovic2014sparsity} transforms the approximation of the linear operator into an optimization problem with constraints in the eigenvalues, modes, or mode amplitudes. Multi-resolution DMD \cite{kutz2016multiresolution} and higher-order DMD \cite{le2017higher} provide a recursive way to improve the frequency resolution and transient handling capability of the standard DMD. Kernel-based DMD \cite{kevrekidis2015kernel} and Extended DMD (EDMD) \cite{williams2015data} provide a means to extend the framework towards nonlinear systems by creating meaningful observables. Proctor et al. \cite{proctor2016dynamic} developed a variant of DMD known as DMD with controls (DMDc) to incorporate the input signals into the DMD framework. Benner et al. \cite{Benner2018} developed an extension of DMDc, known as input-output DMD (ioDMD), by providing means to incorporate outputs alongside inputs and states. The ioDMD and DMDc algorithm provide an elegant way to extend the standard DMD to include the effects of external forcing functions.

Although DMD is widely popularised across diverse fields such as fluid flow \cite{kaiser2018sparse, yuan2021flow, rowley2009spectral, Benner2018}, 
epidemiology 
\cite{10.1093/inthealth/ihv009},  neuroscience \cite{brunton2016extracting}, video processing \cite{grosek2014dynamic}, {to our knowledge 
it has not been widely employed in the field of structural dynamics.}
This gap in the literature is partly attributed to the strong dominance of principal component analysis among structural dynamics researchers \cite{yan2005structural, krishnan2018real} and also to the requirement of high dimensional data (full state measurements) for DMD \cite{kutz2016dynamic}. The requirement of full state measurements for the DMD is somewhat restrictive, limiting the methodology's application in fields with high dimensional data. In practice, for mechanical systems, responses can only be measured at specific strategic locations owing to limitations in the acquisition hardware, and internal full model operators are seldom available.
For low dimensional systems and systems with limited measurements, researchers have taken inspiration from Taken's embedding theory \cite{takens1981detecting} and proposed applying the DMD procedure on  time-shifted coordinates \cite{2158-2491_2014_2_391, brunton2016extracting, kramer2017sparse, yuan2021flow, le2017higher}.  These methods require accurate tuning of the time delays (or hyper-parameters), which is often problem-specific, and noise in the observed data may lead to erroneous results  \cite{erichson2019randomized, yuan2021flow}. Recent works have proposed techniques to address these problems and the interested reader is referred to, e.g.,  \cite{kramer2017sparse,kaiser2018sparse,uy2021active, goyal2021learning} and the references therein for more details. 

{In the present study, for systems with a limited number of measurements a novel data-driven methodology called Wavelet-based Dynamic mode Decomposition (WDMD) is proposed.
The proposed methodology builds on  ioDMD and utilizes the wavelet decomposition of measured responses as observables, thereby enlarging the state dimensions. 
In other words, wavelet coefficients of the measured outputs serve as the pseudo-states of the dynamical system, and the DMD framework approximates a linear operator that advances the pseudo-states by a time step. The present approach can be thought of as a particular case of EDMD with a special choice of the observables. 
Hence, the present methodology can be applied to model dynamical systems such as a vibrating mechanical system, whose internal state vectors are not readily available, and only data from a few output locations are accessible, which is often the case in practice. }

In the present context, data-driven modeling creates numerical models for capturing the 
input-output characteristics of the underlying structure. The advantage with such models is that it relies completely on measured responses, thereby circumventing the need for the knowledge of any underlying governing dynamics of the structure.
In this paper, the data corresponds to time-domain samples of the input-output trajectories, and modeling equates to the best fit linear operator that advances the system states by a time step. 
In this paper, the data is fixed, only some outputs are observed and one cannot go back to re-query the dynamics. The authors refer the reader to \cite{uy2021active, uy2021operator}, which deals with the partial state observation where it is possible to re-query the  system dynamics at each stage.
For data-driven techniques that uses frequency domain samples, the authors refer the reader to \cite{AntBG20,malladi2021estimating, loewnerchapter, krishnan2022leveraging,mayo2007framework, drmac2015vector,gustavsen1999rational} and the references therein.

The major contributions of this papers are as follows: First, a new data-driven modeling methodology based on WDMD that utilizes only the input-output trajectories of the system is proposed. To the authors’ best knowledge, extending ioDMD’s applicability through the use of wavelets has not yet been explored. The proposed methodology provides at least the same and in many cases better quality of the fit using only input-output trajectories, compared with the baseline approach, which has access to full state information.
In addition, 
our present work provides a means to apply the DMD algorithm towards modal analysis and structural vibration of a mechanical system.
 Finally, these numerical results are complemented
with experimental tests on a free-free aluminum cantilever beam, in which WDMD methodology is utilized to develop data-driven models from (noisy) real data. 

The paper is structured as follows. First, a brief description of DMD, ioDMD, and EDMD are presented in \Cref{sec1}. In \Cref{sec2}, our work's major contribution, WDMD, is derived. The proposed methodology is demonstrated on data from a simulated finite element model of a cantilever beam in \Cref{numexp}. In \Cref{expstudty},  experimental case studies are carried out on a free-free beam to demonstrate the efficiency and robustness of the WDMD in approximating practical mechanical systems. Conclusions and potential future directions are given in \Cref{conc}.

\section{Background}\label{sec1}
\subsection{Dynamic mode decomposition}
In this section, a brief introduction to the classical dynamic mode decomposition (DMD) framework is provided. For details, we refer the reader to \cite{schmid_2010, dmdbook,mauroy2020koopman} and the references therein. Consider the system of time-invariant ordinary differential equations of the form
\begin{equation}\label{eq1}
\dot{\mathbf{x}}(t) = f(\mathbf{x}(t)),\
\end{equation}
where  $x(t) \in {\Re ^N}$ is the state vector and $f:{\Re ^N} \to {\Re ^N}$ is a nonlinear map. 
Given the  sampling times $t_0,t_1,\ldots,t_K$ (equally distanced), let 
$\{\mathbf{x}(t_0),~\mathbf{x}(t_1),~\ldots,~\mathbf{x}(t_K)\}$ denote the samples of the state $\mathbf{x}(t)$ of dynamical system \cref{eq1}. For this data, define two snapshot matrices 
${{\mathbf{X}}_0}$ and ${{\mathbf{X}}_1}$ as
\begin{equation} \label{statemat}
{{\mathbf{X}}_0} = 
\begin{bmatrix}
{\mathbf{x}}({t_0}) & {\mathbf{x}}({t_1}) & \cdots &
{\mathbf{x}}({t_{K-1}})
\end{bmatrix} \in {\Re ^{N \times K}}
~~\mbox{and}~~
{{\mathbf{X}}_1} = 
\begin{bmatrix}
{\mathbf{x}}({t_1}) & {\mathbf{x}}({t_2}) & \cdots &
{\mathbf{x}}({t_{K}})
\end{bmatrix}\in {\Re ^{N \times K}},
\end{equation}
where ${{\mathbf{X}}_0}$ denotes the snapshot matrix from $t_0$ to $t_{K-1}$ and ${{\mathbf{X}}_1} \in {\Re ^{N \times K - 1}}$  from $t_1$ to $t_K$, which advances the ${{\mathbf{X}}_0}$ matrix by one time step.
The most fundamental form of DMD aims to explain the snapshot data with a  linear dynamical  system of the form,
\begin{equation}\label{eq2}
\mathbf{x} ({t_{k+1}}) \approx \mathbf {A}\mathbf{x} ({t_k}),~~
\mbox{for}~~k=0,1,2,..,K-1,~\mbox{where}~\mathbf{A} \in {\Re ^{N \times N}}.
\end{equation}
In terms of ${{\mathbf{X}}_0}$ and ${{\mathbf{X}}_1}$,
the approximation in \cref{eq2} can be written in the matrix form as 
\begin{equation}
    \mathbf{X}_1\approx \mathbf{A}\mathbf{X}_0.
\end{equation}
  The DMD algorithm finds the best-fit solution $\mathbf{A}$, one that minimizes the least-squares distance in the Frobenius norm, i.e.,         
\begin{equation}\label{eq4}
\mathbf{A} = \underset{\hat{\mathbf{A}} \in \mathbb{R}^{N \times N}}{\arg \min }\left\|\mathbf{X}_{1}-\hat{\mathbf{A}} \mathbf{X}_0\right\|_{F}.
\end{equation}
The optimal solution $\mathbf{A}$ in  \cref{eq4} is given by
\begin{equation}\label{eq5}
{\mathbf{A}} = {{\mathbf{X}}_1}{{\mathbf{X}}_0}^ \dagger,
\end{equation}
where ${\mathbf{X}_0}^ \dagger \in {\Re ^{K \times N}}$ denotes the Moore-Penrose inverse of  ${\mathbf{X}_0} \in {\Re ^{N \times K}}$. 
The practical issues in computing the $\mathbf{A}$ matrix involves algebraic assumptions and singular value decomposition of the $\mathbf{X}_0$ matrix, which are skipped here for brevity. Interested readers are directed to \cite{dmdbook,alla2017nonlinear,mauroy2020koopman,drmac2019data,drmavc2020dynamic} for more rigorous discussion on practical algorithms and computational considerations.

\subsection{input-output Dynamic mode decomposition (ioDMD)}\label{iodmdwork}

The DMD method as described in the previous section  can only be used for systems that evolve on their own, with no external input and for which all states are assumed to be measured. However, dynamical systems in practice have external inputs and the form as represented by \cref{eq2} will not be sufficient to explain the dynamics \cite{kou2019dynamic}. This lead to the development of DMD with controls (DMDc) \cite{proctor2016dynamic} by including measurements of a control input $\mathbf{u}(t)$. The input-output DMD is a further extension of the DMDc by incorporating the observed outputs \cite{annoni2016wind, Benner2018}. The ioDMD framework constructs a reduced-order model directly from the observed input-output data and the full state vector $\mathbf{x}(t)$. The ioDMD framework models the dynamical systems of the form
\begin{equation}\label{sys}
\begin{aligned} \dot{\mathbf{x}}(t) &=f(\mathbf{x}(t), \mathbf{u}(t)), \\ \mathbf{y}(t) &=g(\mathbf{x}(t), \mathbf{u}(t)), \end{aligned}  
\end{equation}
where $\mathbf{u}(t) \in {\Re ^{m}}$ denotes the inputs driving the system and $\mathbf{y}(t)\in {\Re ^{d}}$ is the measured output.

The ioDMD method approximates the evolution of \cref{sys} with a linear dymnamical system of the form
\begin{equation}\label{dsc}
\begin{aligned} \mathbf{x}(t_{k+1}) &\approx \mathbf{A} \mathbf{x}(t_{k})+\mathbf{B} \mathbf{u}(t_{k}), \\ \mathbf{y}(t_{k}) &\approx \mathbf{C} \mathbf{x}(t_{k})+\mathbf{D} \mathbf{u}(t_{k}), 
\end{aligned}
\end{equation}
where ${\mathbf{A}} \in {\Re ^{N \times N}},{\mathbf{B}} \in {\Re ^{N \times m}},{\mathbf{C}} \in {\Re ^{d \times N}},{\mathbf{D}} \in {\Re ^{d \times m}}$. In addition to the state snapshots 
$\mathbf{X}_0$ and $\mathbf{X}_1$ in \cref{statemat}, define the input and output snapshot matrices as
\begin{equation}\label{u0y0}
\begin{array}{l}
{\mathbf{U_0}} = \left[ {\begin{array}{*{20}{c}}
{{\mathbf{u}}({t_0})}&{{\mathbf{u}}({t_1})}&... &{{\mathbf{u}}({t_{K-1}})}
\end{array}} \right] \in {\Re ^{m \times K}}
~~\mbox{and}~~
{{\mathbf{Y}}_0} = \left[ {\begin{array}{*{20}{c}}
{{\mathbf{y}}({t_0})}&{{\mathbf{y}}({t_1})}&... &{{\mathbf{y}}({t_{K - 1}})} 
\end{array}} \right]\in {\Re ^{d \times K}}.\\
\end{array}
\end{equation}
 This would entail writing \cref{dsc} in terms of its matrix counterpart as
\begin{equation}\label{orgsol}
\left[\begin{array}{c}\mathbf{X}_{1} \\ \mathbf{Y_0}\end{array}\right]\approx\left[\begin{array}{ll}\mathbf{A} & \mathbf{B} \\ \mathbf{C} & \mathbf{D}\end{array}\right]\left[\begin{array}{l}\mathbf{X_0} \\ \mathbf{U_0}\end{array}\right].
\end{equation}
Let
\begin{equation}\label{Geq}
   \fullG=\left[\begin{array}{ll}\mathbf{A} & \mathbf{B} \\ \mathbf{C} & \mathbf{D}\end{array}\right] \in \Re^{(N+d) \times(N+m)}
\end{equation}
denote the optimal solution to \cref{orgsol}.
Similar to the original DMD framework, the ioDMD method finds the optimal solution $\fullG$  by solving a least-squares problem, namely
\begin{equation}\label{minGeq}
  \fullG=  \underset{\hat{\fullG} \in \Re^{(N+d) \times(N+m)}}{\arg \min }\left\|~ \left[\begin{array}{c}\mathbf{X_1} \\ \mathbf{Y_0}\end{array}\right]-\hat{\fullG}\left[\begin{array}{l}\mathbf{X_0} \\ \mathbf{U_0}\end{array}\right]~\right\|_{\mathrm{F}}.
\end{equation}
The optimal $\fullG$ in ioDMD is given by
\begin{equation}\label{equationinv}
    \fullG=\left[\begin{array}{cc}\mathbf{A} & \mathbf{B} \\ \mathbf{C} & \mathbf{D}\end{array}\right] 
    =\left[\begin{array}{c}\mathbf{X_1} \\ \mathbf{Y_0}\end{array}\right]\left[\begin{array}{c}\mathbf{X_0} \\ \mathbf{U_0}\end{array}\right]^{\dagger}.
\end{equation}
Practical issues that arise in computing \cref{eq5} arise here as well.  Computing the pseudo-inverse in \cref{equationinv} often involves inverting small non-zero singular values, thereby leading to numerical instabilities. Therefore, in practice,  singular values below a relative tolerance $\beta \in \Re^{+}$are truncated during the pseudo-inverse computation or other regularization techniques could be employed. {One can also perform  model reduction on the state snapshot matrix $\mathbf{X}$ to further reduce the state-space dimension of the output linear dynamical system. This is carried out by performing an additional SVD-based projection step before solving the least squares problem} in \cref{minGeq}. We refer the reader to \cite{dmdbook,Benner2018} for details. 

\subsection{Extended DMD and the Koopman operator}
Koopman theory \cite{koopman1931hamiltonian} has received considerable attention recently due to the pioneering work of Mezi\'{c} et al. \cite{mezic2005spectral}. It has been shown that the DMD algorithm is a special case of Koopman theory applied to linearly consistent data \cite{rowley2009spectral, mezic2005spectral} and the DMD modes approximate the Koopman eigenvalues if 
the set of observable is sufficiently large (i.e., it  spans the eigenvectors of the Koopman operator)  and {the data has to be sufficiently rich (i.e., it  covers the dynamics of interest)}
\cite{dmdbook, williams2015data}.

In the case of a linear system with full measurements, linear observable or full state measurements are sufficient as they span the eigenvectors of the Koopman operator.
However, in the case of partially observed state measurement or non-linear systems, direct application of the DMD algorithm falls short of recovering the underlying dynamics. 
This observation has lead to the development of extended DMD (EDMD) by Williams et al.~\cite{williams2015data}, {which creates new observables, $\bm{\obs}$, from the state vector.} To give context to the proposed algorithm, WDMD, it is pertinent to introduce EDMD, and the present section serves to do so. The Koopman operator $\mathcal{K}$ acts directly on observables $\bm{\obs}$  rather than on state-space  \cite{koopman1931hamiltonian, williams2015data}, i.e.,
\begin{equation}
\mathcal{K} \bm{\obs} \triangleq \bm{\obs} \circ \mathcal{F} \quad \Rightarrow \quad \mathcal{K} \bm{\obs}\left(\mathbf{x}(t_k)\right)=\bm{\obs}\left(\mathbf{x}(t_{k+1})\right),
\end{equation}
where $\mathcal{F}: \mathcal{M} \rightarrow \mathcal{M}$ is the evolution operator, and $\circ$ denotes the composition operator.  
Intuitively, the linear Koopman operator takes a scalar function $\obs$ and returns a new function $\mathcal{K} \obs$ that predicts the value of $\obs$, one step ahead in future. 
It is to be noted that the dynamical system defined by $\mathcal{F}$ and the one defined by $\mathcal{K}$ are two different parameterizations of the same fundamental behavior.

EDMD aims to  approximate the Koopman operator using a suitable choice of dictionary of observables, $\mathcal{D}=[\obs_1,\obs_2,...,\obs_{N_K}]$. The vector valued function $\mathbf{\Obs}: \mathcal{M} \rightarrow \mathbb{C}^{1 \times N_K}$, where
\begin{equation}\label{observ_phi}
\mathbf{\Obs}(\boldsymbol{x})=[\obs_1(\boldsymbol{x}),\obs_2(\boldsymbol{x}),...,\obs_{N_K}(\boldsymbol{x})],
\end{equation}
can now be defined for the snapshot of the system, $\mathbf{x}(t_k)$, and for  $\mathbf{z}(t_k)=\mathcal{F}(\mathbf{x}(t_k))$. Then, 
one proceeds by defining two snapshot matrices using the samples
$\mathbf{\Obs}(\mathbf{x}(t_k))$
 and $\mathbf{\Obs}(\mathcal{F}(\mathbf{x}(t_k)))$ for $k=0,1,\ldots,K$
 and solves a least-squares problem to form the best fit 
 matrix $\mathbf{K} \in \Re^{N_K \times N_K}$. The eigenvalues and eigenvectors of the finite dimensional representation $\mathbf{K}$ are then used to compute an approximation to the Koopman modes and Koopman eigenfunctions.  Thus, EDMD provides a mean to approximate the infinite dimensional Koopman operator and Koopman eigenfunctions through the selection of observables. For details, the reader is referred to 
\cite{williams2015data}.
\section{Wavelet-based dynamic mode decomposition}\label{sec2}
The main idea behind the present methodology is to create observables using the stationary wavelet coefficients of all the available output measurements, and thereby to approximate the Koopman operator that advances these observables by a time step. Central to this idea is the wavelet transform.
The next subsection, which mainly follows \cite{mallat1999wavelet},
presents a brief overview of  the necessary background on wavelet transform.  We refer the reader to
\cite{daubechies1992ten,mallat1999wavelet,percival2000wavelet} for details. 

\subsection{Maximal overlap discrete wavelet transform (MODWT)}\label{modwtsec}
The wavelet {transform} convolves a signal with a function called the mother wavelet, and the transform is computed across several scales representing different frequency bands for different segments of the signal. The wavelet transform provides a multi-resolution analysis (in contrast to the Fourier transform, which has a uniform time-frequency distribution.) The orthogonal wavelet decomposition of a signal $y(t)$ is given by,
\begin{equation}
    y(t)=\sum_{j} \sum_{k} w_{k}^{j} \psi_{k}^{j},
\end{equation}
with the wavelet coefficient $w_k^j$  given by the inner product 
\begin{equation}\label{dwt}
\begin{aligned}
w_k^j(y) = &
\left\langle y, \psi_k^{j} \right \rangle = 
\left\langle {y(t),\frac{1}{{{2^{j/2}}}}\psi \left( {\frac{t}{{{2^j}}} - k} \right)} \right\rangle = \frac{1}{{{2^{j/2}}}}\int_{ - \infty }^\infty  {y(t)} {\psi ^*}\left( {\frac{t}{{{2^j}}} - k} \right)dt, 
\end{aligned}\
\end{equation}
where the function $\psi$ represents the mother wavelet, $\psi_k^{j} = \frac{1}{2^{j/2}}{\psi}\left( {\frac{t}{{{2^j}}} - k} \right)$ is the
scaled and translated 
mother wavelet, and $(\cdot)^*$ denotes the complex conjugation.
The transform is usually computed at discrete values in a grid corresponding to dyadic values of $2^j$ and translations of $k$, where both $j,k$ are integers, yielding the discrete wavelet transform (DWT). 

In practice, successive high and low-pass filtering replaces the integration procedures in \cref{dwt}. This is followed by down-sampling at each level
\cite{mallat1999wavelet}. The coefficients resulting from these operations are called approximation and detail coefficients.
The details of DWT implementation are not mentioned here for brevity, and interested readers are referred to seminal works such as \cite{mallat1999wavelet}.
Due to down sampling at each level, DWT wavelet coefficients do not have the property of time invariance.

The aforementioned issue can be addressed through a special type of wavelet transform known as the maximal overlap discrete wavelet transform (MODWT) \cite{percival2000wavelet}. 
MODWT has the advantage that it can eliminate down-sampling, thereby resulting in wavelet detail and scale coefficients at each level of the same length as the original time series, thereby facilitating a ready
comparison between the series and its decomposition. 
Decomposing the time-series $\{y(t_0), y(t_1),...,y(t_{K-1})\}$  using MODWT to $J$ levels involves the application of $J$ pairs of filters. The filtering procedure at $j^{th}$ level entails applying a
high-pass filter ($\widetilde{h}^0_{j, l}$) known as wavelet filter, and low-pass filter ($\widetilde{g}^0_{j, l}$) known as scaling filter, where $l=1,2,..L_j$ is the length of the filter. This procedure yields a set of wavelet and scaling coefficients at each level $j$ as
\begin{equation}\label{fil}
    \begin{aligned}
\widetilde{W}_{j, t_k} &=\sum_{l=0}^{K-1} \widetilde{h}_{j, l} y(t_{k-l \ mod \ K}) ,
   ~~~~
\widetilde{V}_{j, t_k} &=\sum_{l=0}^{K-1} \widetilde{g}_{j, l} y(t_{k-l  \ mod \ K}) ,
\end{aligned}
\end{equation}
where $\widetilde{h}_{j, l}$ is $\widetilde{h}^0_{j, l}$ periodized to length $K$ and  $\widetilde{g}_{j, l}$ follows analogously using $\widetilde{g}^0_{j, l}$ values, and mod represents the modular operator; see \cite{percival2000wavelet} for details. The equivalent wavelet filter ($\widetilde{h}_{j, l}$) and scaling filter ($\widetilde{g}_{j, l}$) for the $j^{th}$ level are a set of
scale-dependent localized differencing and averaging operators, respectively, and can be regarded as stretched versions of the base filter ($j=1$). The MODWT wavelet coefficients at each scale will have the same length as the original signal $y(t)$ as seen from \cref{fil}.
Define the time-series vector \begin{equation} \label{eq:ys}
\boldsymbol{y}=[y(t_0), y(t_1),...,y(t_{K-1})]^T \in \Re^{K}.\end{equation} 
Then \cref{fil} can  be expressed in matrix form as
\begin{equation}
    \widetilde{\mathbf{W}}_{j}=\widetilde{\mathbf{\mathcal{W}}}_{j} \ys
    ~~\mbox{and}~~
    \widetilde{\mathbf{V}}_{j}=\widetilde{\mathbf{\mathcal{V}}}_{j} \ys,
\end{equation}
where 
{$\widetilde{\mathbf{W}}_{j} 
= [\widetilde{W}_{j,t_0},~\widetilde{W}_{j,t_1},~\ldots, \widetilde{W}_{j,t_{K-1}}]^T
\in \Re^{K}$ and $\widetilde{\mathbf{V}}_{j}
= [\widetilde{V}_{j,t_0},~\widetilde{V}_{j,t_1},~\ldots, \widetilde{V}_{j,t_{K-1}}]^T
\in \Re^{K}$} represent the $j^{th}$ level MODWT wavelet and scaling coefficients, respectively. The $K \times K$ matrix $\widetilde{\mathbf{\mathcal{W}}}_j$ is defined as
\begin{equation}\label{wamat}
\widetilde{\mathbf{\mathcal{W}}}_j =\frac{1}{2^{k}}\begin{bmatrix}
\widetilde{h}_{j,0} \  & \widetilde{h}_{j,K-1} \  & \widetilde{h}_{j,K-2} \  & \dotsc  & \widetilde{h}_{j,3} & \widetilde{h}_{j,2} & \widetilde{h}_{j,1}\\
\widetilde{h}_{j,1} & \widetilde{h}_{j,0} & \widetilde{h}_{j,K-1} & \dotsc  & \widetilde{h}_{j,4} & \widetilde{h}_{j,3} & \widetilde{h}_{j,2}\\
\widetilde{h}_{j,2} & \widetilde{h}_{j,1} & \widetilde{h}_{j,0} & \dotsc  & \widetilde{h}_{j,5} & \widetilde{h}_{j,4} & \widetilde{h}_{j,3}\\
\vdots  & \vdots  & \vdots  & \dotsc  & \vdots  & \vdots  & \vdots \\
\widetilde{h}_{j,K-2} & \widetilde{h}_{j,K-3} & \widetilde{h}_{j,K-4} & \dotsc  & \widetilde{h}_{j,1} & \widetilde{h}_{j,0} & \widetilde{h}_{j,K-1}\\
\widetilde{h}_{j,K-1} & \widetilde{h}_{j,K-2} & \widetilde{h}_{j,K-3} & \dotsc  & \widetilde{h}_{j,2} & \widetilde{h}_{j,1} & \widetilde{h}_{j,0}
\end{bmatrix},\\
\end{equation}
and the $K \times K$ matrix $\widetilde{\mathbf{\mathcal{V}}}_j$ is
defined analogously using $\widetilde{g}_{j, l}$ values; see \cite{percival2000wavelet} for details. The original time series $y$ can be recovered from its MODWT via
\begin{equation} \label{modwtdecomp}
    \ys=\sum_{j=1}^{J}
    \widetilde{\mathbf{\mathcal{W}}}_j^{\mathrm{T}} \widetilde{\mathbf{W}}_{j}+\widetilde{\mathbf{\mathcal{V}}}_j^{\mathrm{T}} \widetilde{\mathbf{V}}_{j}=\sum_{j=1}^{J} \widetilde{\mathbf{D}}_{j}+\widetilde{\mathbf{S}}_{J}~~~\mbox{where}~~~\widetilde{\mathbf{D}}_{j} := \widetilde{\mathbf{\mathcal{W}}}_j^{\mathrm{T}} \widetilde{\mathbf{W}}_{j}\in\Re^K~~~\mbox{and}~~~\widetilde{\mathbf{S}}_{j}=\widetilde{\mathbf{\mathcal{V}}}_j^{\mathrm{T}} \widetilde{\mathbf{V}}_{j}\in\Re^K.
\end{equation}
The last equality defines a MODWT-based multi-resolution analysis (MRA) of the original time series $\ys$ {in terms of $j^{th}$ level MODWT detail coefficients
$\widetilde{\mathbf{D}}_{j}$ and $j^{th}$ level MODWT smooth coefficients $\widetilde{\mathbf{S}}_{j}$.}

\subsection{Main approach}
Consider an underlying dynamical system evolving in an $N$-dimensional state-space, i.e.,
\begin{equation}\label{sys2}
 \dot{\mathbf{x}}(t) =f(\mathbf{x}(t), \mathbf{u}(t)),
\end{equation}
where $\mathbf{x}(t) \in \Re^{N}$ is the state, $\mathbf{u}(t) \in \Re^{m}$ is  the input, and $f: \Re^N \to \Re^N$ is a nonlinear mapping. Assume that, unlike in DMD or ioDMD, we do not have access to the full-state samples $\mathbf{x}(t_k)$. Instead, 
we have only access to a measurement vector (output)  $\mathbf{y}(t) \in \Re^d$ via an observation (output) matrix $\mathbf{C} \in \Re^{d \times N}$, i.e., we have access to the output 
\begin{equation} \label{out2}
\mathbf{y}(t) = \mathbf{C} \mathbf{x}(t).
\end{equation}
Assume that dynamics are sampled at time instances $t_0,t_1,\ldots,t_K$, yielding the measurement samples
\begin{equation}\label{otptvec}
    \mathbf{y}(t_k)=\mathbf{C}\mathbf{x}(t_k),~~~\mbox{for}~~~k=0,1,\ldots,K-1.
\end{equation}
Based on MODWT analysis of the previous section, our goal is now to create new auxiliary state variables (and an observation matrix) so that the new dynamics with the auxiliary state still corresponds to the true output samples in \cref{otptvec}. Then we can apply the ioDMD using the trajectories of the original input and outputs, and  the trajectories of the auxiliary states. 

Towards this goal, let $y_i(t)$, for $i={1,2,. . .,d}$, denote the $i$th component (row) of the measurement vector $\mathbf{y}(t)$, i.e., $y_i(t)$ is the $i$th output. Decompose $y_i(t)$ using MODWT as in \cref{modwtdecomp}:
\begin{equation} \label{yidecomps}
    \ys_i=\sum_{j=1}^{J} \widetilde{\mathbf{D}}_{j}^{(i)}+\widetilde{\mathbf{S}}_{J}^{(i)} \qquad \mbox{where} \qquad
    \boldsymbol{y}_i=[y_i(t_0), y_i(t_1),...,y_i(t_{K-1})]^T,
\end{equation}
and {${\mathbf{D}}_{j}^{(i)} \in \Re^K$ and ${\mathbf{S}}_{j}^{(i)} \in \Re^K$  are the corresponding $j$th level detail and smooth coefficients corresponding to $y_i(t)$.}
Let $\mathbf{e}_{k} \in \Re^K$ denote the $k$th canonical vector and 
$\mathbf{e} = [1 \ 1 . . .\ 1]^T \in \Re^{J+1}$ denote the vector of ones. 
Then, using \cref{yidecomps}, $y_i(t_k)$ (the $(K+1)$st row of $\ys_i=[y_i(t_0), y_i(t_1),...,y_i(t_{K-1})]^T$) can be written as 
\begin{equation}\label{iwt}
y_i(t_k) = \mathbf{e}_{k+1}^T \ys_i = 
\sum_{j=1}^{J} \mathbf{e}_{k+1}^T \widetilde{\mathbf{D}}_{j}^{(i)}+ \mathbf{e}_{k+1}^T \widetilde{\mathbf{S}}_{J}^{(i)} = \mathbf{e}^T\textcolor{black}{{\mathbf{w}_i(t_k)}}
\end{equation}
where
\begin{equation}
\mathbf{w}_{i}( t_{k}) = \begin{bmatrix}
\mathbf{e}_{k+1}^{T}\widetilde{\mathbf{D}}_{1}^{( i)} \ \\
\mathbf{e}_{k+1}^{T}\widetilde{\mathbf{D}}_{2}^{( i)} \ \\
\vdots \\
\mathbf{e}_{k+1}^{T}\widetilde{\mathbf{D}}_{J}^{( i)} \ \\
\mathbf{e}_{k+1}^{T}\widetilde{\mathbf{S}}_{J}^{( i)} \ 
\end{bmatrix} \in \Re^{J+1}.
\end{equation}
 Then, the full output vector at time $t_k$, i.e., $\mathbf{y}(t_k)$ in \cref{otptvec}, can be rewritten as
\begin{equation}
    \mathbf{y}( t_{k}) =\begin{bmatrix}
y_{1}( t_{k})\\
y_{2}( t_{k})\\
\vdots\\
y_{d}( t_{k})
\end{bmatrix} =\begin{bmatrix}
\mathbf{e}^T\mathbf{w}_{1}( t_{k})\\
\mathbf{e}^T\mathbf{w}_{2}( t_{k})\\
\vdots\\
\mathbf{e}^T\mathbf{w}_{d}( t_{k})
\end{bmatrix}.
\end{equation}
Using the last formula, we define the new auxiliary state $\mathbf{z}(t)$
and  the observation matrix $\mathbf{C_{w}}$
\begin{gather} \label{definez}
\mathbf{z}( t_{k}) =\begin{bmatrix}
\mathbf{w}_{1}( t_{k}) \\
\mathbf{w}_{2}( t_{k}) \\
\vdots\\
\mathbf{w}_{d}( t_{k})
\end{bmatrix} \in \Re^{d(J+1) } 
~~\mbox{and}~~
\mathbf{C_{w}} =\begin{bmatrix}
\mathbf{e}^{T} & 0 & \cdots & 0 & 0\\
0 & \mathbf{e}^{T} & \ddots & 0 & 0\\
\vdots & \ddots & \ddots & \ddots & \vdots\\
0 & 0 & \ddots & \mathbf{e}^{T} & 0\\
0 & 0 & \cdots & 0 & \mathbf{e}^{T}
\end{bmatrix} \in \Re^{d \times d(J+1)},
\end{gather}
so that 
\begin{equation}\label{mapwave}
    \mathbf{y}(t_k)=\mathbf{C_w}\mathbf{z}( t_{k}).
\end{equation}
Note that  the new auxiliary state variable $\mathbf{z}(t)$ 
is composed of the wavelet coefficient observables and thus its samples encodes how the  wavelet coefficient observables evolve over time. Moreover, with the observation matrix $\mathbf{C_{w}}$,  the true output/measurement vector $\mathbf{y}(t)$ is written in terms of the new state variable $\mathbf{z}(t)$.

Given \emph{only} the output snapshots $\mathbf{y}(t_k)$ in \cref{otptvec} of the underlying dynamical system, use \cref{yidecomps}-\cref{definez} to construct the snapshot matrix $\mathbf{Z}$ of the wavelet coefficient observables as 
\begin{gather}  \label{zwdmd}
\mathbf{Z} =[\mathbf{z}( t_{0}) \ \ \mathbf{z}( t_{1}) \ \ \mathbf{z}( t_{2}) \ \ .\ .\ .\ \mathbf{z}( t_{K})] \in 
\Re^{d(J+1)\times (K+1)}. 
\end{gather}
Also construct the input snapshot matrix $\mathbf{U}_0$ and 
$\mathbf{Y}_0$
\begin{equation}\label{u0y0wdmd}
\begin{array}{l}
{\mathbf{U_0}} = \left[ {\begin{array}{*{20}{c}}
{{\mathbf{u}}({t_0})}&{{\mathbf{u}}({t_1})}&... &{{\mathbf{u}}({t_{K-1}})}
\end{array}} \right] \in {\Re ^{M \times K}}
~~\mbox{and}~~
{{\mathbf{Y}}_0} = \left[ {\begin{array}{*{20}{c}}
{{\mathbf{y}}({t_0})}&{{\mathbf{y}}({t_1})}&... &{{\mathbf{y}}({t_{K - 1}})} 
\end{array}} \right]\in {\Re ^{d \times K}}.\\
\end{array}
\end{equation}
Note that while the input snapshot matrix 
$\mathbf{U}_0$ and 
the output snapshot matrix $\mathbf{Y}_0$ 
in \cref{u0y0wdmd}
correspond to the true inputs and outputs of the underlying dynamical system \cref{sys2} and \cref{out2}, 
the state snapshot matrix $\mathbf{Z}$ in \cref{zwdmd}
are obtained via the wavelet coefficient observables (as the original state measurements $\mathbf{x}(t_k)$ are not available). Then, WDMD represents the snapshot triplets $\mathbf{Z},\mathbf{U}_0$ and $\mathbf{Y}_0$ with  the   dynamical system 
    \begin{equation}  \label{wdmdsys}
    \mathbf{z}(t_{k+1}) \approx \mathbf{A_w} (t_k) \mathbf{z}(t_k) + \mathbf{B_w u}(t_k),~~~
    \mathbf{y}(t_k) \approx \mathbf{C_w} \mathbf{z}(t_k) 
    + \mathbf{D_w} \mathbf{u}(t_k).
    \end{equation}
    This reformulation of the input/output data via wavelet coefficient observables to input/state/output data
    allows to apply ioDMD  to construct the 
    matrices $\mathbf{A_w}$, $\mathbf{B_w}$, $\mathbf{C_w}$,  and $\mathbf{D_w}$ via a least-squares fit as in
    \Cref{iodmdwork}. Towards this goal,
define the two matrices    
\begin{align}  \label{eq:Z0Z1}
    \mathbf{Z}_{0} &= [\mathbf{z}(t_0),~\mathbf{z}(t_1),\ldots,\mathbf{z}(t_{K-1})] \in {\Re ^{d.(J+1) \times K}}
~~\mbox{and}~~
    ~~\mathbf{Z}_{1} = [\mathbf{z}(t_1),~\mathbf{z}(t_1),\ldots,\mathbf{z}(t_K)] \in {\Re ^{d.(J+1) \times K}}.
\end{align}
Then, the dynamical system coefficients in \cref{wdmdsys} 
are given by
 \begin{equation}\label{wveeqn}
    \left[ \begin{array}{cc}{\mathbf A_w} & {\mathbf B_w} \\ {\mathbf C_w} & {\mathbf D_w}\end{array}\right]=\left[ \begin{array}{l}{\mathbf Z_{1}} \\ {\mathbf{Y}_{0}}\end{array}\right] \left[ \begin{array}{l}{\mathbf Z_{0}} \\ {\mathbf U_{0}}\end{array}\right]^{^ \dagger}.
\end{equation}
 A brief algorithmic sketch of WDMD is given in Algorithm \ref{Algorithm}.

\begin{algorithm}[htp] 
 \caption{WDMD algorithm}   
\label{Algorithm}                                            
\algorithmicrequire~ Output measurements $\{\mathbf{y}(t_i)\} \in \Re^d $ and input measurements $\{\mathbf{u}(t_i)\} \in \Re^m$ for
$i=0,1,\ldots,K$.

\algorithmicensure~ State-space model:   $\mathbf{A_w} \in \Re^{d.(J+1) \times d.(J+1)},  \mathbf{B_w}\in \Re^{d.(J+1) \times m}, \mathbf{C_w} \in \Re^{d \times d.(J+1)}$, and $\mathbf{D_w} \in \Re^{d \times m}$.

\begin{algorithmic} [1]                                        
\STATE \label{computeUL}  Using $\{\mathbf{y}(t_i)\}$, construct the wavelet observable snapshots $\{\mathbf{z}(t_i)\}$ using \cref{yidecomps} -- \cref{definez}.
\STATE  \label{svdstep} Form  $\mathbf{Z} =[\mathbf{z}( t_{0}) \ \ \mathbf{z}( t_{1}) \ \ \mathbf{z}( t_{2}) \ \ .\ .\ .\ \mathbf{z}( t_{K})]$ as in \cref{zwdmd}
\STATE \label{WrVrstep}  
 Assemble  $\mathbf{Z_0}$ and $\mathbf{Z_1}$ 
 as in \cref{eq:Z0Z1},  and $\mathbf{Y_0}$ and $\mathbf{U_0}$ as in 
 \eqref{u0y0wdmd}.
	\STATE \label{lasteq}  Compute the approximate discrete linear state-space matrices as in \cref{wveeqn}
	
\end{algorithmic}
\end{algorithm}


Given only the output samples, the present work's major contribution is  to enlarge the original subspace via wavelet decomposition of the response measurement $\mathbf{y}(t)$ and creating new states of the system using 
the wavelet coefficients. 
Therefore, the  WDMD methodology can be considered a special case of EDMD with choice of observables in \cref{observ_phi} resulting from the wavelet coefficients $w_k^j(\boldsymbol{y}) = \langle \boldsymbol{y}, \psi_k^j \rangle$.
Thus,  WDMD provides a set of basis functions or a kernel operator, which lifts the output measurements to wavelet states, thereby potentially spanning the eigenvectors of the Koopman operator
and approximating the Koopman operator through ioDMD. 

\section{A numerical case study using a finite element beam}\label{numexp}
\begin{figure}[!htb]
\centering
\includegraphics[trim=0cm 0cm 0cm 0cm,clip,scale=1]{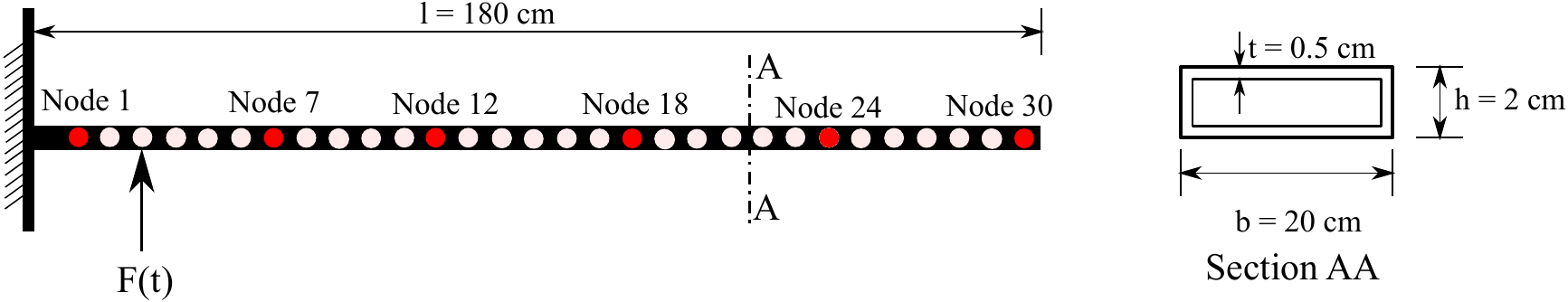}
\caption{Cantilever beam model used for the Finite Element simulations}
\label{femdiag}
\end{figure}
  
Numerical simulations are carried out on a hollow cantilever beam model with the dimensions shown in \Cref{femdiag}. The beam under study is a finite element representation of an Euler-Bernoulli beam with $30$ nodal points representing $60$ degrees of freedom (DOF). Taking the displacement and velocity of each DOF as the states yields a state-space representation in the first-order form

\begin{gather}\label{csstat}
\dot{\mathbf{x}}(t) \ =\ \underbrace{\begin{bmatrix}
\mathbf{0} & \mathbf{I}\\
-\mathbf{M}^{-1}\mathbf{K} & -\mathbf{M}^{-1}
\mathbf{G}
\end{bmatrix}}_{:=\varmathbb{A}}\mathbf{x}(t) \ +\ \underbrace{\begin{bmatrix}
\mathbf{0}\\
-\mathbf{M}^{-1}\mathbf{F}
\end{bmatrix}}_{:=\varmathbb{B}}\mathbf{u}(t),\qquad
\mathbf{y}( t) \ =\ \varmathbb{C}\mathbf{x}(t),
\end{gather}
where $\mathbf{M}, \mathbf{K}, \mathbf{G} \in \Re^{60\times60} $
are, respectively, the mass, stiffness, and damping matrices and $\mathbf{F} \in \Re^{60\times1}$ is the loading vector;  $\mathbf{x} \in \Re^{120 \times 1}$ is the state vector; $\mathbf{u}(t) \in \Re$ is the scalar input; and $\mathbf{y}\in \Re^d$ is the $d$-dimensional output vector. This yields the first-order state-space quantities 
$\varmathbb{A}\in \Re^{120 \times 120}$, $\varmathbb{B}\in \Re^{120}$, and
$\varmathbb{C}\in \Re^{d \times 120}$. 
The outputs, observed in $\mathbf{y}$, can be either the displacement or velocity of the observed nodal points. The choice of output will be further clarified below.

The proposed WDMD approach will be utilized to generate a single-input/multiple-output (SIMO), data-driven approximation to the the beam model in \cref{csstat} using \emph{only} the simulated input-output response of the beam \emph{without} access to its state-space matrices. This model will then be used to simulate the transient dynamic response of the structure to a given excitation (a testing signal) to illustrate the quality of the fit. In addition to this time-domain error measure, the input-output mapping of the data-driven model can also be assessed in the frequency domain by computing the frequency response function (\frf) of the learned model and comparing it with the original \frf. The \frf~of the beam model \cref{csstat}, denoted by $\Hf(\omega)$, is given by
\begin{equation} \label{Hr_in_ss}
\Hf(\omega) = \varmathbb{C}(\imath \omega  \mathbf{I} -\varmathbb{A})^{-1}\varmathbb{B},
\end{equation}
where $\imath^2 = -1$.
Let $\tilde{\mathcal{H}}$ denote the \frf~of the learned model and $\tilde{\mathbf{y}}(t)$ the output of the learned. Then, the following two relative error metrics are defined to evaluate the quality of the fit,
\begin{equation}\label{errmet}
    \epsilon ^{rel}_{fd}=\sqrt{\frac{\sum_{j=1}^{L_{\omega}} \| \mathcal{H}\left( \omega_{j}\right)-\tilde{\mathcal{H}}\left( \omega_{j}\right) \|_{2}^{2}}{\sum_{j=1}^{L_{\omega}}\left\|\mathcal{H}\left( \omega_{j}\right)\right\|_{2}^{2}}}~~\text{and}~~
\epsilon ^{rel}_{td} =\sqrt{\frac{\sum\nolimits ^{K}_{i=1} ||\mathbf{y}( t_{i}) -\tilde{\mathbf{y}}( t_{i}) ||^{2}_{2}}{\sum ^{K}_{i=1}\mathbf{||y}( t_{i}) ||^{2}_{2}}},
\end{equation}
where $L_{\omega}$ and $K$ are the number of frequency samples and time points respectively. While $\epsilon ^{rel}_{fd}$ is the relative error between the original \frf~($\Hf(\omega)$) and the fitted \frf~($\tilde{\Hf}(\omega)$), $\epsilon ^{rel}_{td}$  measures the relative error in  time domain between the measured responses $\mathbf{y}(t)$ and the predicted responses $\tilde{\mathbf{y}}(t)$.

In the present numerical case study, the FEM beam is excited using a chirp input over the frequency range $10 - 800$ Hz. The responses are collected at a sampling frequency of $5000$ Hz. The application of the ioDMD methodology, which assumes access to full state observation, for modeling the input-output response of the FEM beam, is demonstrated in \Cref{ioDMDsec}. Next, in \Cref{wdmdsec} we compare these results with WDMD. In \Cref{compsec} we present a brief comparative study between WDMD and Delay-DMD. As mentioned earlier, one can perform an additional model reduction via an SVD-based projection on the state data to further reduce the learned system dynamics \cite{dmdbook, Benner2018} . {In the present study,  this additional step has yielded  negligible changes to the final data-driven model and thus is skipped in all  the results.}
\subsection{Data-driven modeling using ioDMD}\label{ioDMDsec}
As discussed in \Cref{iodmdwork}, the ioDMD methodology assumes knowledge about the system's full internal states $\mathbf{x}(t)$. Hence, the ioDMD is ideally suited towards grey box modeling wherein the internal states of the system are also sampled. For the beam's finite element model, the internal states represent the displacement and velocity at each degree of freedom. The training package provided to the algorithm consists of: i) the input forcing signal used to excite the structure (chirp signal), ii) the internal state measurements, and iii) the measured output responses. 
In the current example, the ioDMD has access to all the internal states of the system and the output is assumed to be measured at the $6$ nodal points shown in \Cref{femdiag}. The measured displacements at nodes $1,7,12,18,24,$ and $30$ are designated as the output responses in the present section. From the provided training package, the ioDMD algorithm, as presented in \Cref{{iodmdwork}},  generates a linear discrete dynamical system of the form,
\begin{equation}\label{ss}
\begin{aligned} \mathbf{x}(t_{k+1}) &=\mathbf{A} \mathbf{x}(t_{k})+\mathbf{B} \mathbf{u}(t_{k}), \\ \mathbf{y}(t_{k}) &=\mathbf{C} \mathbf{x}(t_{k})+\mathbf{D} \mathbf{u}(t_{k}), 
\end{aligned}
\end{equation}
to approximate the original beam dynamics in \cref{csstat}.
The singular value truncation tolerance in the computation of the pseudoinverse in \cref{equationinv} is set to $\beta=10^{-12}$.
Since no model reduction step is applied to further reduce the system dimension, the learned model's state dimension is equal to the total number of degrees of freedom in the finite element model, i.e., $120$. Since, the output is  measured at $6$ nodal points, the  SIMO ioDMD state-space model in \cref{ss} has $120$ internal states, single input, and six output, and thus the state-space matrices are  $\mathbf{A} \in  {\Re^{120 \times 120}}$, $\mathbf{B} \in  {\Re^{120 \times 1}}$, $\mathbf{C} \in  {\Re^{6 \times 120}}$, and $\mathbf{D} \in  {\Re^{6 \times 1}}$. 

\subsubsection{ioDMD model training and testing results}
\begin{figure}[!h]
\centering
\includegraphics[trim=1cm 0cm 0cm 0cm,scale=0.9]{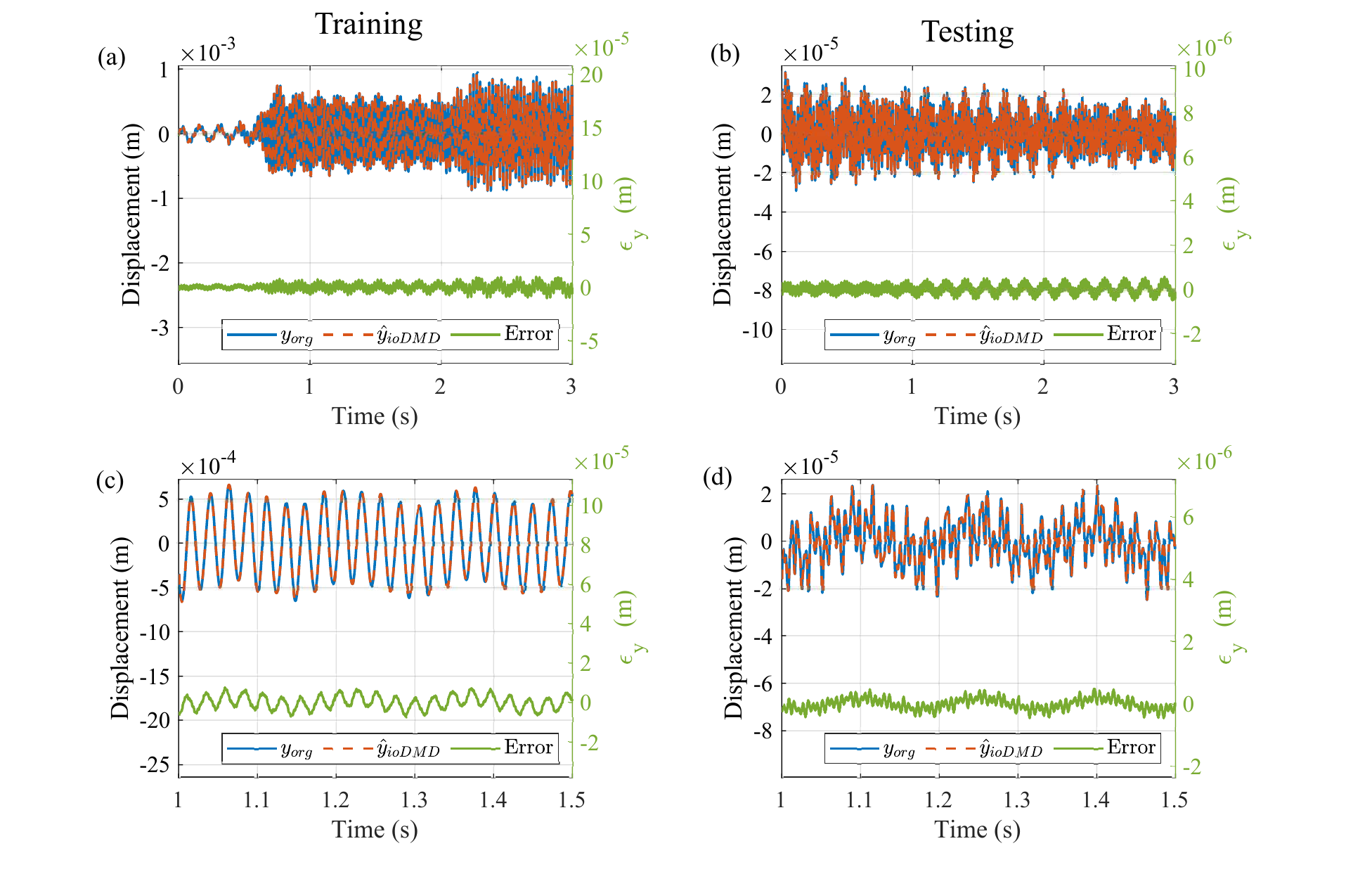}
\caption{(a) Comparison of the predicted response and the original response measured at node $18$ when excited with the same training signal (chirp); (b) Comparison of the predicted response and the original response measured at node $18$ when excited with a sine burst at $165$ Hz (testing data); (c) Zoomed up version of (a) demonstrating a high-fidelity approximation in the training phase (d) Zoomed up version of (b) demonstrating a high-fidelity approximation in the testing phase}
\label{ioDMD}
\end{figure} 
The results of modeling the dynamic response of the FEM beam using ioDMD are summarized in \Cref{ioDMD}. The data-driven ioDMD model is excited using the same chirp signal used for training the model. For demonstration purposes, among the six outputs, the predicted response output at node $18$ ($\hat{y}$) is compared with the measured output from the FEM simulations ($y$)  and is shown in \Cref{ioDMD}(a). The low value of the time domain error ($\epsilon_y = y - \hat{y}$ ), shown in green, in \Cref{ioDMD}(a) demonstrates the good quality of the fit. The relative error of $\epsilon_{td}^{rel} = 2.3\times 10^{-2}$ further substantiates the good quality of the fit across all the $6$ outputs in the time domain. For validation purposes, a sine burst at $165.1$ Hz is used to excite the ioDMD model and quality of the fit analyzed. The \Cref{ioDMD}(b) contrasts the simulated response from the ioDMD methodology with the original response measurement from node $18$ in the testing case. The low value of the error plot in \Cref{ioDMD}(b) clearly illustrates the validity of the model over the frequency ranges of interest. \Cref{ioDMD}(c) and \Cref{ioDMD}(d) presents zoomed versions of the training and testing case respectively. The relative time domain error for ioDMD 
is  {$\epsilon_{td} ^{rel} = 1.62 \times 10^{-2}$} for the testing case. 
\begin{figure}[!htb]
\centering
\includegraphics[trim=0cm 0cm 0cm 0cm,scale=0.8]{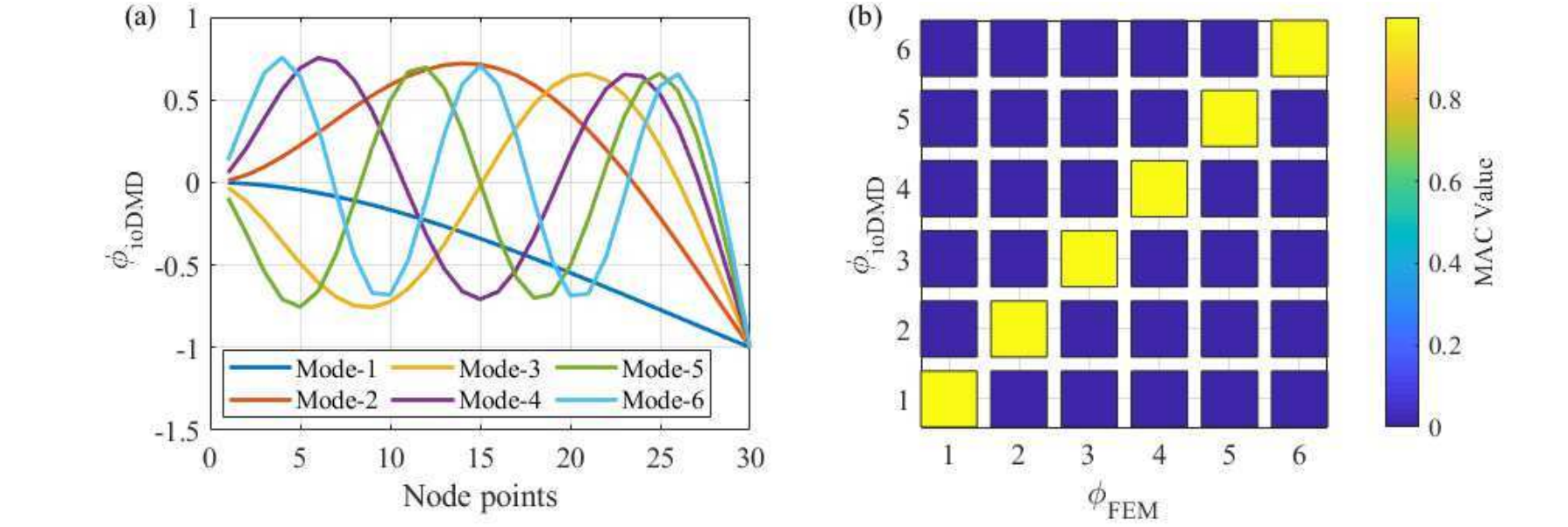}
\caption{(a) Mode shapes extracted by ioDMD (b) Comparison of the ioDMD modeshapes with the analytical mode shapes using modal assurance criterior (MAC)}
\label{modeshapes_mac}
\end{figure}

{It is straightforward to recover the dynamic modes of the system under consideration using the developed state-space model}. {The finite element model is setup in such a way that the dynamic modes of the beam corresponds to the modes of vibration of the system} \cite{albakri2020estimating}. The recovered modes ($\mathbf{\phi}_{ioDMD})$, as shown in \Cref{modeshapes_mac}(a), closely resemble the modes of vibration of a cantilever beam ($\mathbf{\phi}_{FEM})$. The quality of the modes recovered using the ioDMD methodology is evaluated using the model assurance criteria (MAC) \cite{allemang2003modal}. If individual columns of $\mathbf{\phi}_{ioDMD}$, representing the DMD modes, are a close match with that of $\mathbf{\phi}_{FEM}$, then the MAC value will be close to $1$. The value of 1 in the diagonal term in \Cref{modeshapes_mac}(b) shows that the recovered DMD correlates well with the actual modes of vibration of the system. The zeros in the off-diagonal position further validate the orthogonality of the DMD modes (also a property of the physical modes), thus further demonstrating the efficacy of the ioDMD model in accurately capturing the hidden dynamics of the system under consideration.
\subsection{Data-driven modeling using WDMD}\label{wdmdsec}
\begin{figure}[!htb]
\centering
\includegraphics[trim=0cm 0cm 0cm 0cm,scale=1]{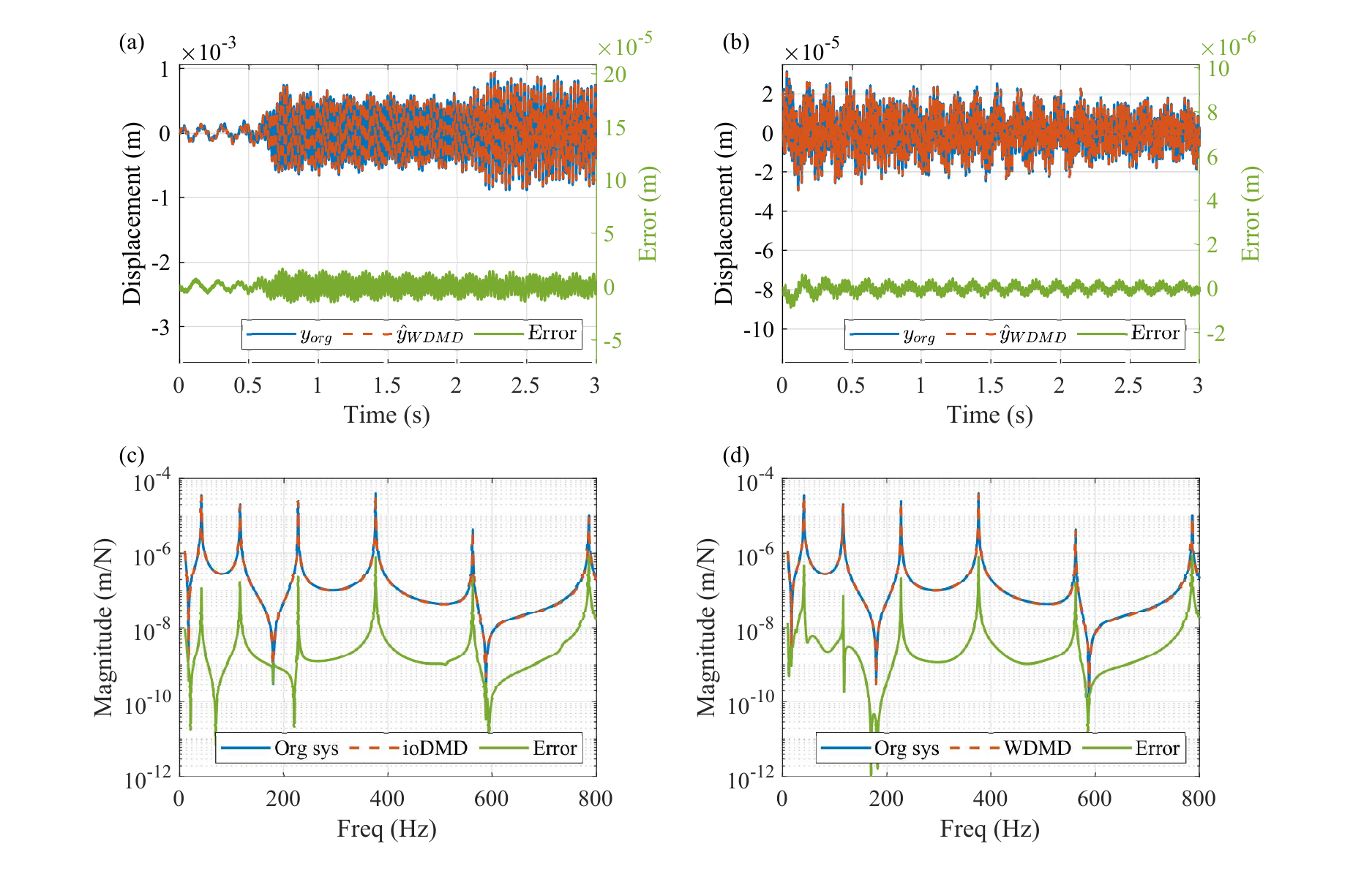}
\caption{(a) Comparison of the predicted response using WDMD and the original response measured at node $18$ when excited with the training input, plotted alongside with the error in green; (b) Comparison of the predicted response using WDMD and the original response measured at node $18$ when excited with the sine burst input at $165.1$ Hz (testing phase), plotted alongside with the error in green; (c) Comparison of the magnitude of analytical and predicted \frf~at node $18$ using ioDMD alongside with error magnitudes; (d) Comparison of the magnitude of analytical and predicted \frf~at node $18$ using WDMD alongside with error magnitudes.}
\label{wdmd}
\end{figure}
\begin{figure}[!htb]
\centering
\includegraphics[trim=0.5cm 0cm 0cm 0cm,scale=0.9]{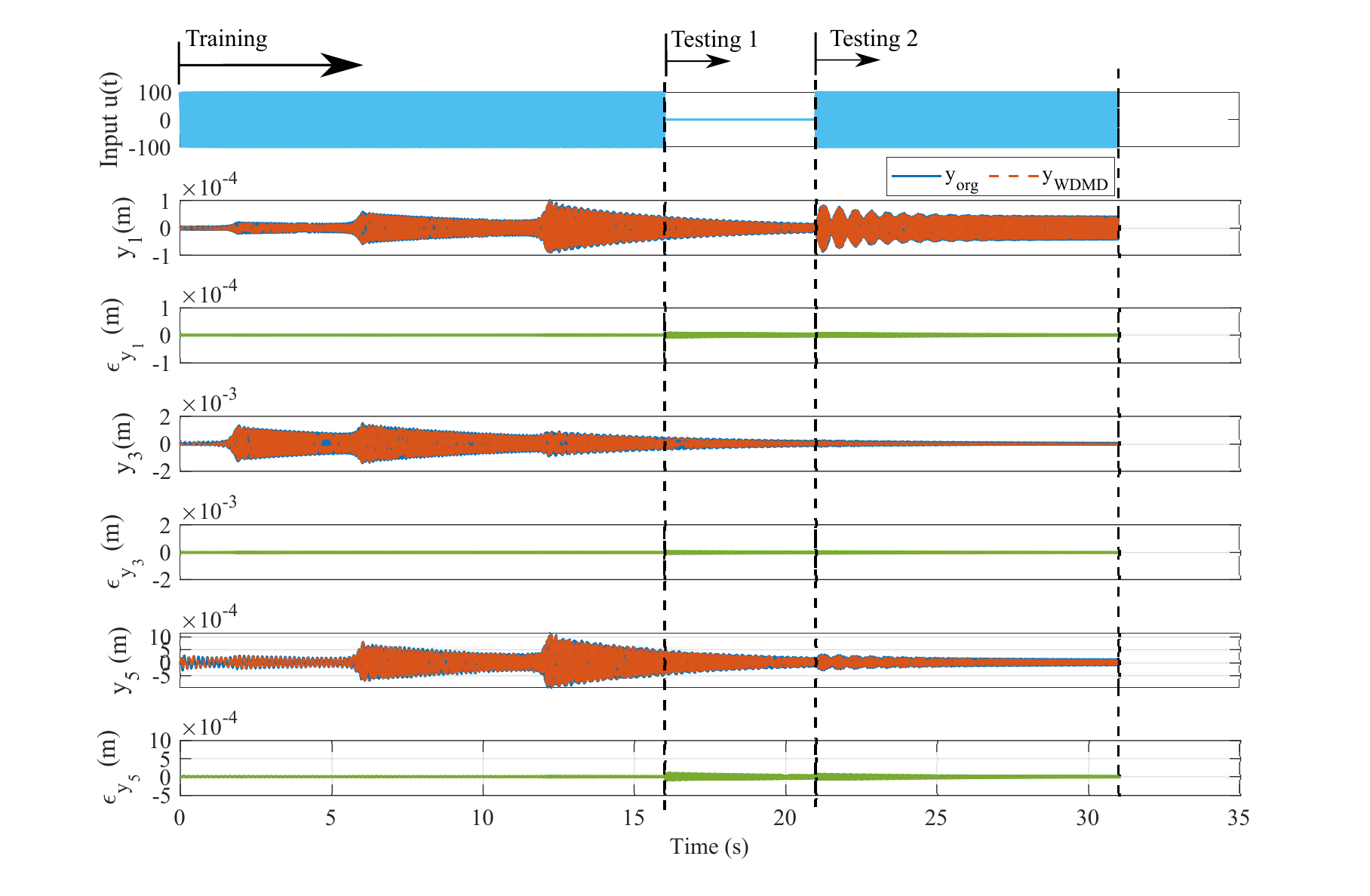}
\caption{Training phase followed by testing phases, comparing the predicted responses from  WDMD model with the FEM simulation results.}
\label{traintest}
\end{figure}
We now apply the WDMD methodology to model the input-output dynamic responses of the simulated beam. WDMD is used to obtain a data-driven model, using the snapshots matrices of only the measured outputs at $d = 6$ locations and the chirp input, recorded in training phase. It is important to note that WDMD  develops a SIMO, data-driven, state-space model by \textit{only} utilizing the input-output trajectories of the measured nodal points, thus circumventing the restrictive assumption  to require the samples of  all the latent states of the system.
{It will be shown in later sections that WDMD yields high-fidelity approximates even with a small number of outputs and the quality of the fit further improves with an increase in number of measured outputs.} The parameters controlling the WDMD algorithm are i) the type of wavelet and ii) the level of wavelet decomposition. In the current study, the Haar wavelet \cite{mallat1999wavelet} is selected as the default setting throughout and the level of decomposition in this section is $J=13$. As in the ioDMD case, the singular value truncation tolerance in computing the pseudoinverse is set to $\beta = 10^{-12}$.
Based on these parameters, the number of auxiliary states in the resulting WDMD model in
\cref{wdmdsys} is given by $d \times (J+1) = 84$. Finally, this results in a  
state-space matrices with the following dimensions  $\mathbf{A_w} \in  {\Re^{84 \times 84}}$, $\mathbf{B_w} \in  {\Re^{84 \times 1}}$, $\mathbf{C_w} \in  {\Re^{6 \times 84}}$, and $\mathbf{D_w} \in  {\Re^{6 \times 1}}$.

\begin{figure}[!htb]
\centering
\includegraphics[trim=0cm 0cm 0cm 0cm,scale=0.8]{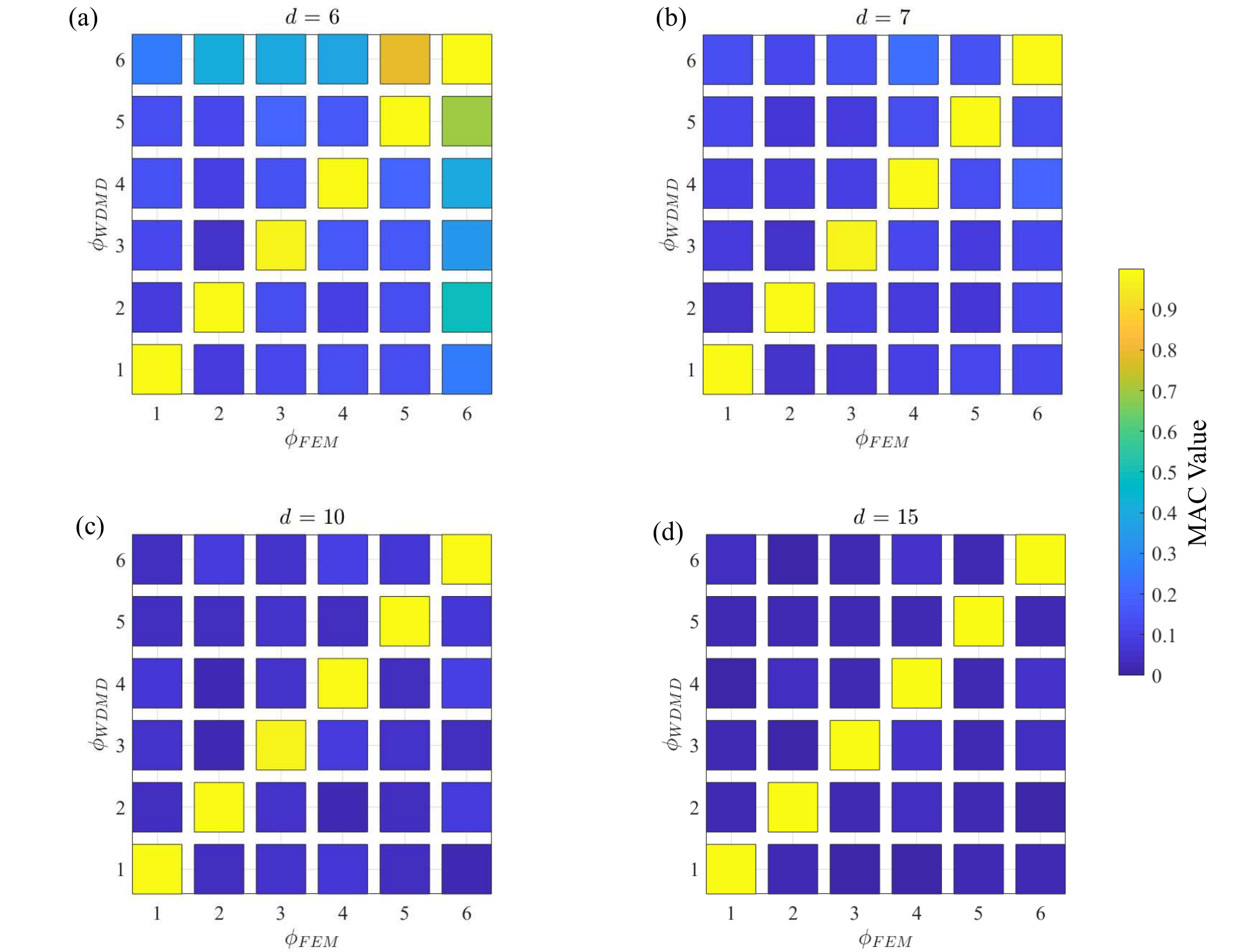}
\caption{MAC comparison plots for WDMD for (a) $d = 6$; (b) $d = 7$; (c) $d = 10$; (d) $d = 15$}
\label{mac_wdmd}
\end{figure}

\subsubsection{WDMD model training and testing results}
The results of modeling the dynamic response of the FEM beam using WDMD are summarized in \Cref{wdmd}. For a first comparison, the data-driven model produced by WDMD is used to reproduce the behavior of the system during the training phase. To be consistent, we excite the WDMD model with the same training (chirp) and testing signal (sine burst) as in the ioDMD case. \Cref{wdmd}(a) shows the predicted response at node $18$ alongside with the measured output from the FEM simulations, illustrating a high-fidelity match. The low value of the time domain error (green) in \Cref{wdmd}(a) further validates the good quality of the fit. The relative time-domain error of {$\epsilon_{td} ^{rel} = 1.59\times 10^{-2}$} across the $6$ predicted response illustrates a better performance over ioDMD in this case. The sine burst at $165.1$ Hz excites the SIMO WDMD model and the predicted response at node $18$ is compared with that of FEM simulation results in \Cref{wdmd}(b). The testing phase results in a relative  error of
{$\epsilon_{td} ^{rel} = 7.84 \times 10^{-3}$.}

To better illustrate the frequency domain performance, 
\Cref{wdmd}(c) and \Cref{wdmd}(d) depict, respectively, the magnitude of the predicted \frf~($\tilde{\Hf}_{iodmd}(\omega)$) due to ioDMD and predicted \frf~($\tilde{\Hf}_{wdmd}(\omega)$) due to WDMD, as compared to the original \frf~($H(\omega)$)  measured at node $18$.  While ioDMD results in a frequency domain relative error of {$\epsilon_{fd}^{rel}=1.51\times 10^{-2}$}, WDMD results in a slightly higher error value of {$\epsilon_{fd}^{rel}=1.63\times 10^{-2}$}. Nevertheless, both WDMD and ioDMD demonstrates excellent capability to capture the frequency domain characteristics of the system. And more importantly WDMD achieves this accuracy by  measuring only $6$ of the state variables out of the total $120$.

To further test the capabilities of the WDMD model, two different testing phases are performed.  The first test (Test $1$) consists of a time interval with no excitation, thereby allowing the system to be driven by the initial conditions at the end of the training phase. The second test (Test $2$) consists of a sine burst at $230.4$ Hz. It is pertinent to observe that at no point during the beginning of a phase,  the input conditions are corrected. This is particularly challenging  for Test 1 where there is no input and {thus small deviations in initial conditions can result in high errors.} \Cref{traintest} shows $3$  out of the $6$ predicted responses alongside with the error between the WDMD model and the original FEM model. WDMD
produces a high-quality fit for the all phases of these tests as can be seen from the low value of the errors, thus illustrating the efficacy of the algorithm.

Similar to the ioDMD case, the quality of the WDMD modes ($\phi_{WDMD})$ and their agreement to the physical modes of the beam ($\phi_{FEM})$ can be examined by using the MAC plots as shown in \Cref{mac_wdmd}, where {$d$} represents the number of outputs measured. 
The quality of the extracted dynamic modes improves with the increase in total number of outputs measured as seen from \Cref{mac_wdmd}. Nevertheless, even with a small number of measured output ($d=6$), WDMD resulted in a data-driven model that was able to meaningfully extract modal characteristics of the leading six modes as seen from the diagonal terms in \Cref{mac_wdmd} (a). At $d = 10$, we see that WDMD results almost converge to the ioDMD MAC plots. 
\begin{figure}[!htb]
\centering
\includegraphics[trim=1.3cm 0cm 0cm 0cm,scale=1.0]{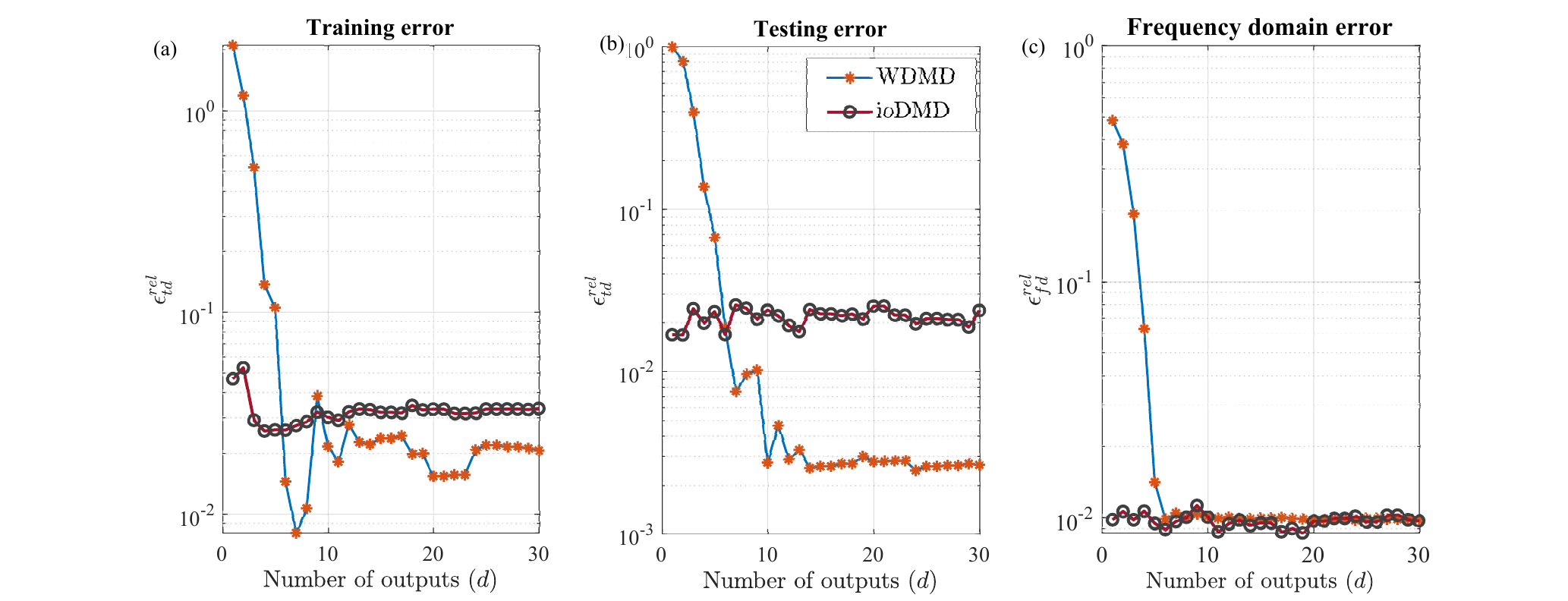}
\caption{Error convergence study results for the SIMO case: (a) Relative time domain error for training phase; (b) Relative time domain error for testing phase; (c) Relative frequency domain error}
\label{conv_simo}
\end{figure}

Since WDMD only relies on input-output trajectories at the observed nodes, the major factor affecting the WDMD methodology's efficacy is the total number of outputs measured across the system. This necessitates an error convergence study in the time domain as well as in the frequency domain. By increasing the number of outputs measured, the quality of the fit improves as seen from \Cref{conv_simo}. The figure shows the relative error as a function of the total number of measured outputs ($d$). The figure also provides the ioDMD model error values for comparison purposes. Even with fewer measurements (as few as $6$), for this example, the WDMD methodology outperforms the ioDMD methodology. 
This figure demonstrates the advantage of applying the WDMD in the practical situation wherein only a handful of the output trajectories can be measured. The same is true for the frequency domain representation. The relative error, $\epsilon_{fd}^{rel}$ also drops as a function of the number of measurements available but eventually converges to the ioDMD model error of around $0.015.$
\subsection{Results for the multiple input case - WDMD}
In this section, a multiple-input multiple-output (MIMO), state-space model using  WDMD is developed. Towards this goal, the FEM beam is subjected to uncorrelated input excitations at node $3$ and node $17$, thereby simulating multiple (two) input excitations.  Similar to the  SIMO case, WDMD builds a data-driven model using measured outputs at selected nodal points and chirp inputs that are used for excitation, resulting in a MIMO learned model 
as in \cref{wdmdsys}. WDMD uses $J=13$ as before. 
\begin{figure}[!htb]
\centering
\includegraphics[trim=1.2cm 0cm 0cm 0cm,scale=1]{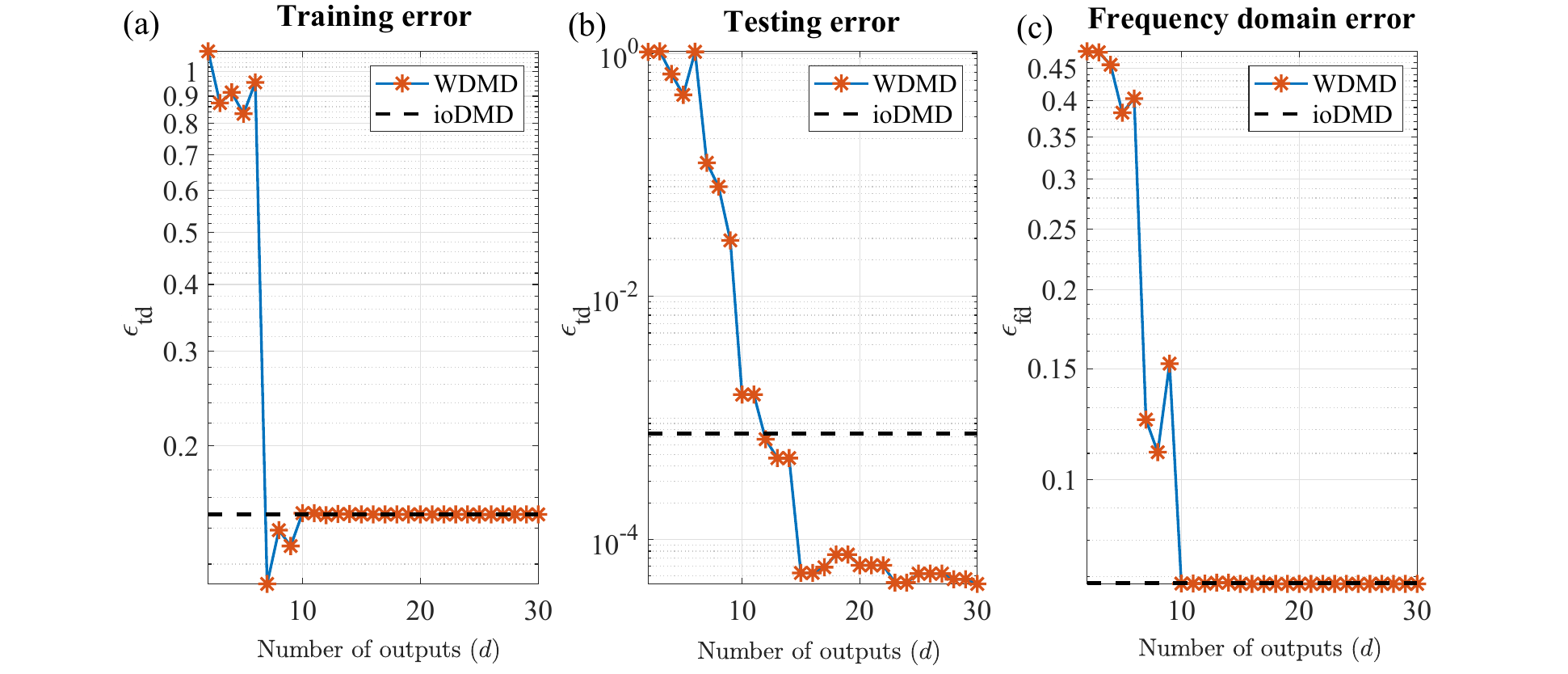}
\caption{Error convergence study results for the MIMO case: (a) Relative time domain error for training phase; (b) Relative time domain error for testing phase; (c) Relative frequency domain error.}
\label{conv_mimo}
\end{figure}
Similar to the SIMO case, error convergence studies for the training and testing cases are performed. \Cref{conv_mimo}(a) and \Cref{conv_mimo}(b) shows error convergence plots for the training and testing case respectively as the number of outputs change. The WDMD model error converges towards the full ioDMD baseline error for the training case, when the number of outputs ($d$) measured are greater than 11. However for the testing case with $d>11$, the WDMD results in a lower testing error compared to the ioDMD model. These results follow the similar pattern to those of the SISO case. The smaller error for WDMD in the testing case for this specific testing input might be due to the frequency content of the signal. 
 \Cref{conv_mimo} depicts the error in the frequency domain. Similar to the time domain error, there is no noticeable reduction in the error beyond $d = 10$ and around this value the WDMD error converges to the ioDMD model error. It is important to emphasize that WDMD provides comparable results to ioDMD (and better in the time-domain testing case for this example) with a much smaller number of observed state. In this example,
 there are a total of $120$ state variables with WDMD observing only $10$  of them (less than $10\%$ of the total). Thus, WDMD is able to match the full-state observation accurately.

\subsection{Comparison with the Delay-DMD}\label{compsec}
\begin{figure}[!htb]
\centering
\includegraphics[trim=0cm 0cm 0cm 0cm,scale=1.0]{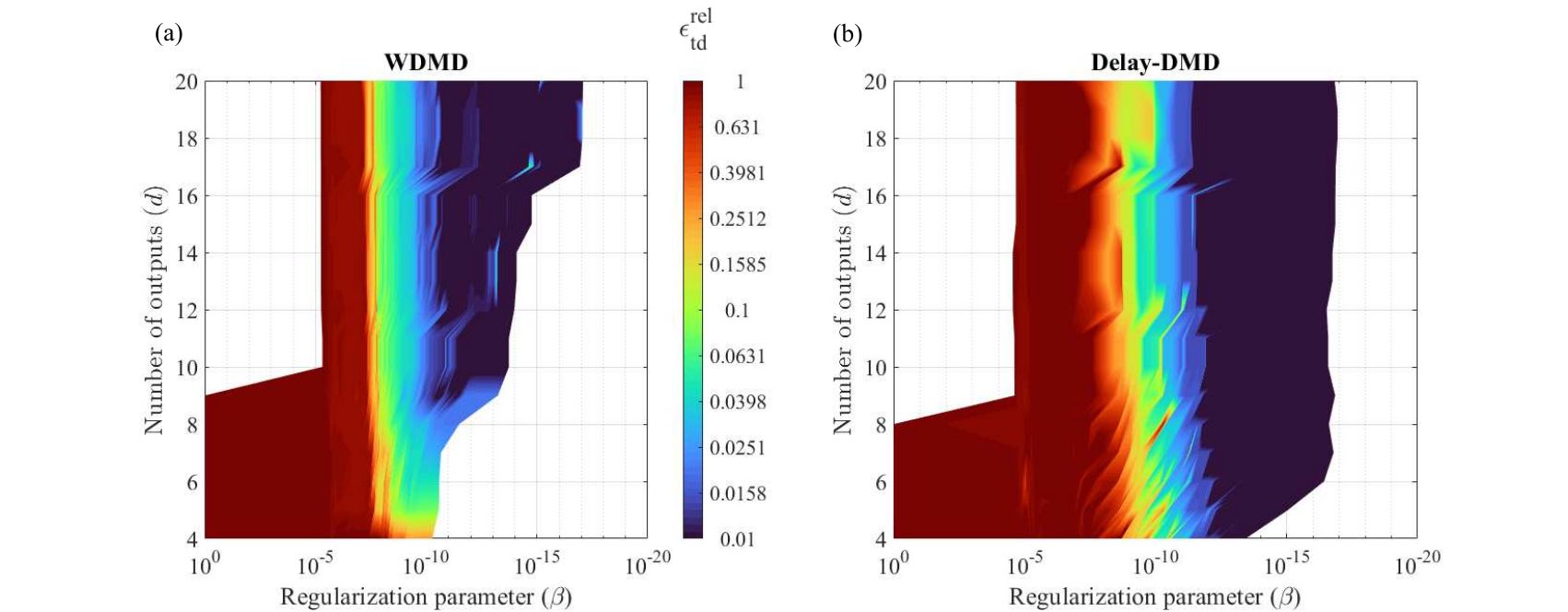}
\caption{Comparison of WDMD with  Delay-DMD  for the noise-free data  using the $\epsilon_{td}^{rel}$ measure as $\beta$ and $d$ vary}
\label{comp}
\end{figure}
In this section, WDMD is compared with  Delay-DMD. Delay-DMD can be thought of as a special case of EDMD with the observable vector in \cref{observ_phi} composed of the time-delayed versions of the measurements as,
 $ \mathbf{\Obs}(\mathbf{x}_k)=[\mathbf{x}(t_k), \mathbf{x}(t_{k-\delta}),...,\mathbf{x}(t_{k-\delta\tau})]$,
where $k = 1,2,...K-\delta\tau$ represent the snapshot indices. The integer $\delta$ and $\tau$ represent the lag-time and the embedding dimension,  respectively. 
As with any algorithm, the performance of the Delay-DMD depends on various parameter choices, such as the lagtime $\delta$ and the embedding dimension $\tau$  that are often problem-specific \cite{kamb2020time,yuan2021flow}. Discussion regarding the parameter choices and embedding dimensions are beyond the scope of the present study. For comparison purposes with WDMD, $\delta$ is chosen as $1$ and the embedding dimension is varied on a per case basis \cite{dmdbook, kamb2020time, yuan2021flow}.

The present comparative study is conducted on the FEM beam model described in \Cref{numexp} with the same training and testing cases. As discussed in {\mbox{\Cref{numexp}}}, the performance WDMD depends on the number of available measurements and the decomposition level. Thus, the number of available measurements is varied as one of the parameters.

The input-output trajectories of the FEM beam model are utilized by WDMD and Delay-DMD towards building a SIMO, data-driven, state-space model. 
For comparison purposes, both WDMD and Delay-DMD model are excited with the same chirp signal to reproduce the system's behaviour during the training phase and the quality of the fit is assessed. Since both of these methods utilize only the measured input-output trajectories, the quality of the fit depends on the total number of measured responses, $d$. 
While WDMD results in a state-space model order of dimension $d(J+1)$, Delay-DMD results in a model order of $d\times\tau$.
Thus, Delay-DMD can lead to higher model orders even with a small number of measured outputs if the embedding dimension $\tau$ is large enough. 

As explained in \Cref{iodmdwork}, {apart from the number of responses measured, the performance of the algorithm also depends on the regularization parameter $\beta$.} Therefore, for every training data set (i.e., for each value of $d$), the parameter $\beta$ is varied in both WDMD and Delay-DMD and the quality of the fit evaluated using $\epsilon_{td}^{rel}$. Therefore, for both WDMD and Delay-DMD the relative error $\epsilon_{td}^{rel}$ is plotted in the form of surface contours with $d$ and $\beta$ being the two axes, as shown in \Cref{comp}(a) and \Cref{comp}(b), respectively.


In \Cref{comp}(a) and \Cref{comp}(b), three distinct regions are observed: i) $\beta  > 10^{-8}$, ii)  $10^{-8} > \beta > 10^{-12}$, and iii)  $\beta < 10^{-12}$. For the first region, both the training and testing errors are large for both methodologies, and this is attributed to the higher value of the regularization parameter, thus leading to an oversimplified model with significant singular values being truncated. However, in the second region, both  methodologies demonstrate better performance, with WDMD having lower training and testing error compared to Delay-DMD. Finally, in the third region, Delay-DMD shows lower training error compared to the WDMD.
\begin{figure}[!htb]
\centering
\includegraphics[trim=0cm 0cm 0cm 0cm,scale=1]{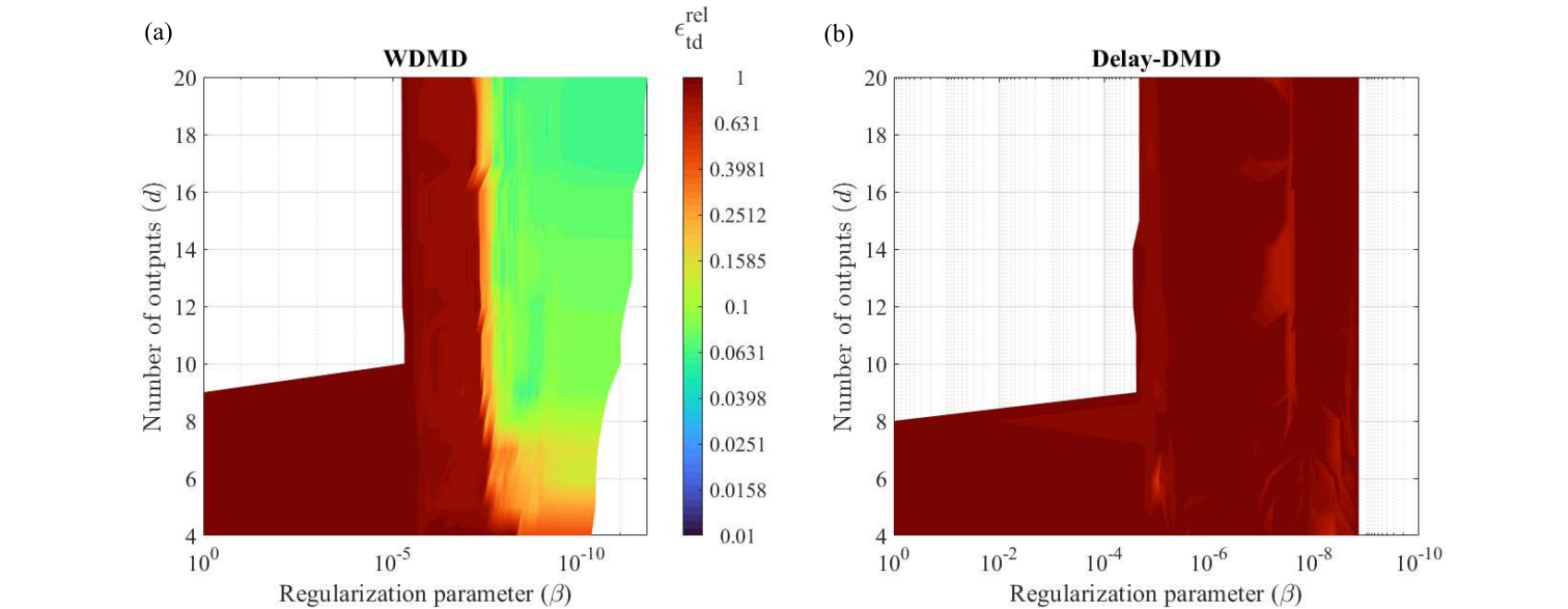}
\caption{Comparison of WDMD with  Delay-DMD  for the $0.5\%$ additive noise case  using the $\epsilon_{td}^{rel}$ measure as $\beta$ and $d$ vary}

\label{comp_noise}
\end{figure}


For most practical situations, the signal obtained from the sensors will be corrupted with noise. This warrants repeating the same set of simulations in the presence of added noise.
Towards this goal, simulations are realized to study the performance of both methodologies in the presence of added noise. Zero-mean Gaussian white noise with  amplitude corresponding to $0.5\%$ of the measured signal is artificially added to all the outputs recorded from the FEM simulations and both methods are repeated. As expected, both methodologies perform poorly for $\beta$ greater than $10^{-5}$ as seen in the surface plots in \Cref{comp_noise}. However, it is observed from \Cref{comp_noise}(a) and \Cref{comp_noise}(b) that for lower values of regularization parameter the training errors in WDMD algorithm are orders of magnitude smaller than Delay-DMD.
{However, by no means this one numerical example claims to show that in the case of noisy data WDMD performs better in general. 
Authors acknowledge the fact that the issues observed with Delay-DMD in this one specific example may be resolved by proper tuning of hyper-parameters, 
even though tuning these parameters for every possible case of noise could be challenging in practice.} \cite{kaiser2018sparse, yuan2021flow}. {The main goal of this numerical study was simply to see how WDMD might behave under noise if it was tuned, i.e., the level of decomposition $J$, for the noise-free data. 
It is possible that the smoothing operation inherent to the wavelet observables used in WDMD might naturally help with the noisy data. However,
a detailed theoretical analysis of WDMD for the noisy case is beyond the scope of this paper and will be done in a future work.}

\section{Experimental study}\label{expstudty}
\begin{figure}[!htb]
\centering
\includegraphics[trim=0cm 0cm 0cm 0cm,scale=1]{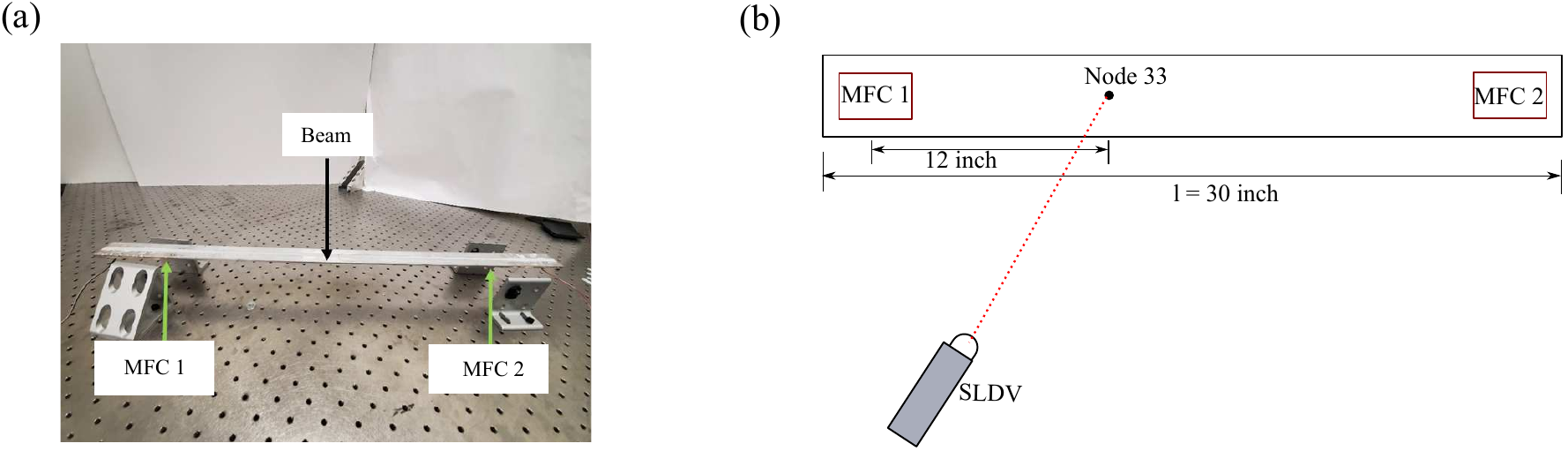}
\caption{Experimental setup used for the study (a) the picture of the free-free beam attached with MFC-1 and MFC-2, (b) schematic displaying the locations of MFC's on the beam and node $33$.}
\label{setup}
\end{figure}

We now present an experimental case study to validate the efficacy of the WDMD in modeling the dynamical response of a beam excited by an external forcing. The experimental set up used to measure the time domain response of the beam is depicted in \Cref{setup}. A $30~in.$ long aluminum beam with a rectangular cross-section of $ 1.5~in. \times 0.1~in.$ (bxh) has been selected for this study. Free-free boundary conditions are approximated by suspending the beam under test with fishing line wires. Two Macro Fiber Composites (MFCs), model number 29K06-005B, are bonded to either end of the beam for excitation purposes. The MFCs are actuated by supplying a Matlab generated signal delivered using an NI DAQ, and amplified through a power amplifier (Trek  PZD350A-2-L). A scanning laser doppler vibrometer (SLDV), Polytec PSV-400, is used to measure the beam's dynamic response when excited with the MFC's. In the present set of experiments, $67$ equally spaced scanning points are defined along the beam's length. The SLDV measures the velocity response of all the scanning points in the beam. The whole assembly has been placed on top of a Newport ST series smart table to isolate the effects of ground vibrations and other random excitations.

Using this setup, two sets of experiments are realized: i) Single-input multiple-output (SIMO) and ii) Multi-input multi-output (MIMO).
Similar to the finite element simulations, the beam under study is excited using two sets of inputs for both of these cases: i) Chirp signal over the frequency range $100$ - $500$ Hz to train the algorithm, and ii) Sine burst to validate the model. The SLDV measures the output response (velocity) at a sampling rate of $5000$ Hz from multiple locations along the beam. Measurement locations are densely selected to have enough data to study the effect of  number of measurement points on the quality of the fit. In this experimental case study, the input corresponds to the voltage supplied to the MFC, while the output  to the measured velocity responses. 
\subsection{Data-driven SIMO model}\label{simo}
In this section, a SIMO, data-driven, state-space model is developed using WDMD  based on the measured responses along the beam, valid over the frequency range of $100$ to $500$ Hz. The WDMD methodology utilizes the measured the velocity responses at $10$ equidistant points along the beam and chirp input voltage supplied to the MFCs for building the data-driven SIMO model. It is pertinent to note that, although we measured velocity at $66$ scanning points in the beam, we utilize only $10$ output responses to develop the data-driven model. The level of wavelet decomposition ($J$) is set as $J=13$. Hence, WDMD outputs a linear discrete state-space form with the following dimensions: $\mathbf{A_W} \in  {\Re^{140 \times 140}}$, $\mathbf{B_w} \in  {\Re^{140 \times 1}}$, $\mathbf{C_w} \in  {\Re^{10 \times 140}}$, and $\mathbf{D_w} \in  {\Re^{10 \times 1}}$. The singular value truncation tolerance $\beta$ is set to $\beta = 10^{-12}$. 
\begin{figure}[!htb]
\centering
\includegraphics[trim=0.8cm 0cm 0cm 0cm,scale=0.95]{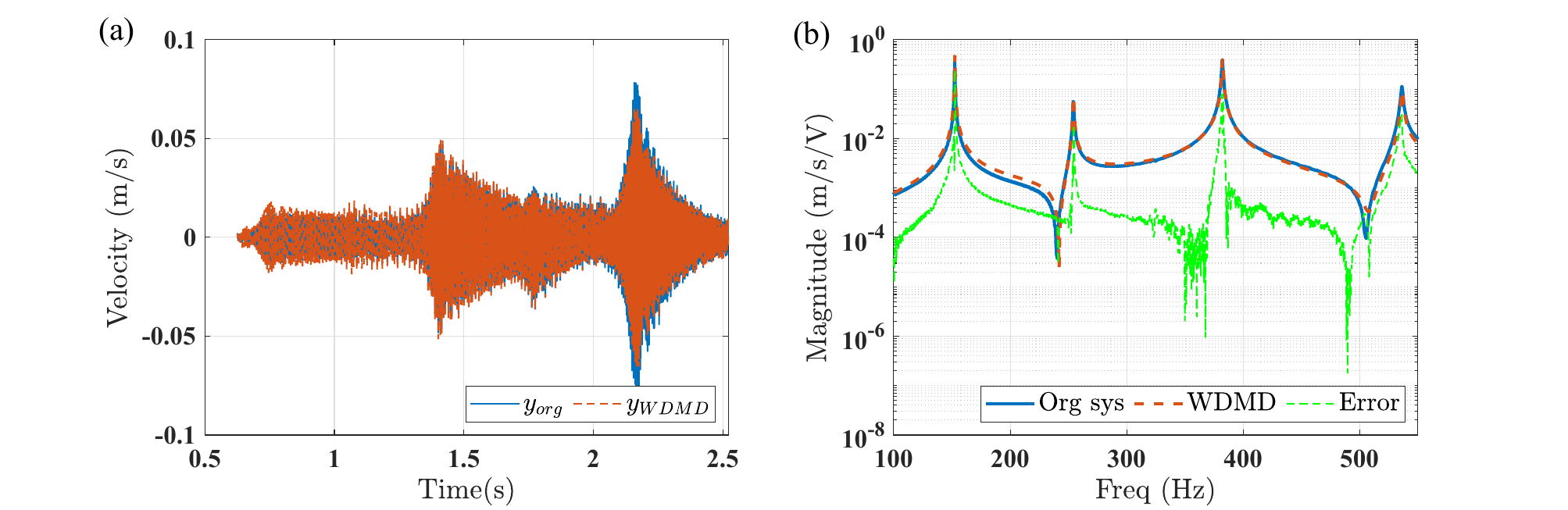}
\caption{Comparison of predicted WDMD response with the actual experimental data in (a) time domain and (b) frequency domain}
\label{33dof}
\end{figure}

Once the data-driven, SIMO, state-space model is developed, the behaviour of the free-free beam during the training phase is reproduced using the recorded chirp input. For demonstration purpose, the predicted response (dashed orange line) at one of the ten measured locations (nodal point $33$) is compared with the SLDV measured velocity (solid blue line) in \Cref{33dof}(a). \Cref{33dof}(b) depicts the magnitude of the predicted \frf~($\tilde{\Hf}_{wdmd}(\omega)$) compared to the experimentally measured \frf~($H(\omega)$) corresponding the same node. The high-fidelity fits in \Cref{33dof}(a) and \Cref{33dof}(b) demonstrate the efficacy of the algorithm in accurately reproducing the time domain and frequency domain characteristics of the beam under test. The relative error 
{$\epsilon_{td}^{rel} = 2.17\times 10^{-1}$} is higher compared to the simulated data cases of the previous section, but this is expected because of the unfiltered experimental noise.
Further studies are needed to further improve the robustness of WDMD methodology in cases with high experimental noise.
It is important to note that employing ioDMD for the present experimental case study is not feasible since the  internal states of the structure under test is unknown.  
\begin{figure}[!htb]
\centering
\includegraphics[trim=0cm 0cm 0cm 0cm,scale=0.75]{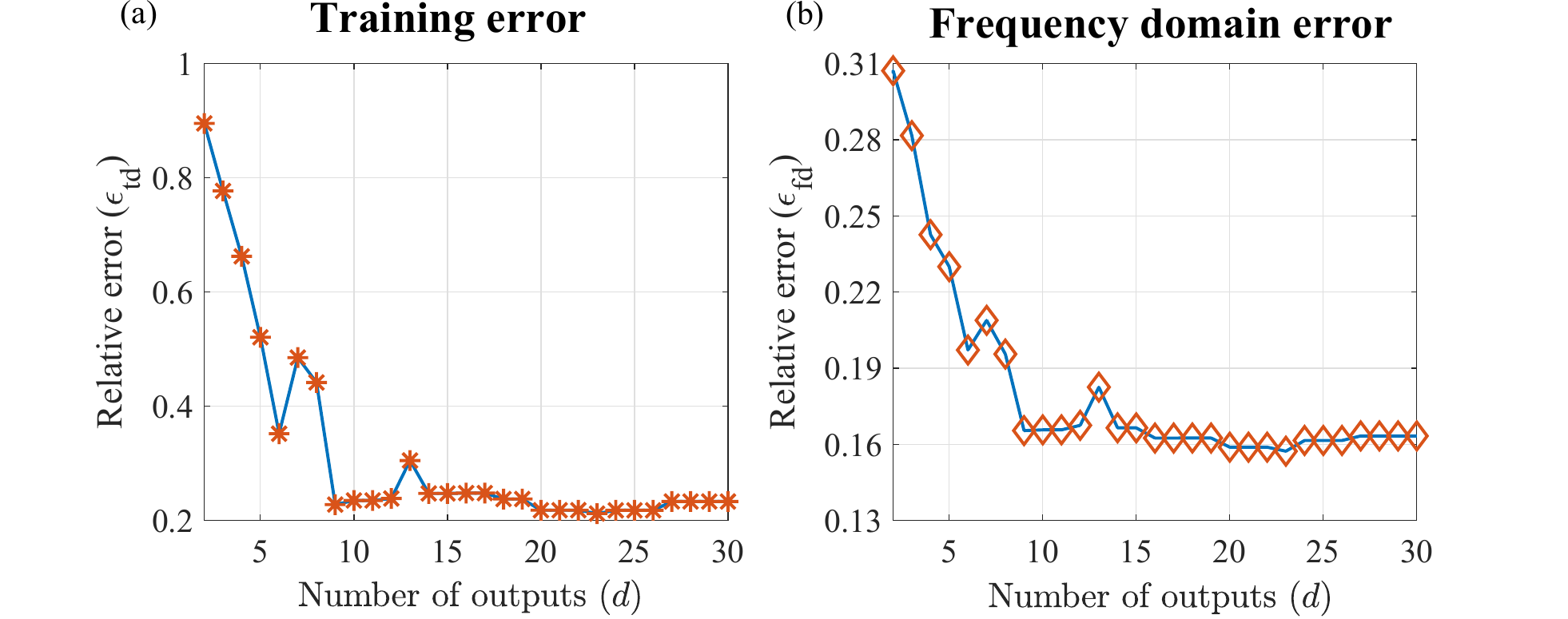}
\caption{Error convergence results for the experimental SIMO case study: (a) Relative time domain error for training phase and (b) Relative frequency domain error}
\label{conv_exp}
\end{figure}

Similar to \Cref{numexp}, we perform error convergence studies to evaluate the quality of the fit as a function of the number of output responses available to the algorithm. 
As before, the data-driven model's quality of the fit is evaluated using $\epsilon_{td}^{rel}$ and $\epsilon_{fd}^{rel}$ in time and frequency domain, respectively. 
The error convergence study is carried out by sequentially varying the number of outputs provided to the WDMD  from $2$ to $30$.
\Cref{conv_exp} shows the error convergence in both time and frequency domains.
The relative error metrics $\epsilon_{td}^{rel}$ and $\epsilon_{fd}^{rel}$ converges at {$2.17\times 10^{-1}$ and $1.64\times 10^{-1}$}, respectively, at $d=9$  and no further significant improvement in the quality of the fit is  observed for  $d \geq 9$.
Nevertheless, the experimental studies clearly show the efficacy of WDMD methodology in accurately modeling the dynamic response of a beam. 
\subsection{Data-driven MIMO model}
The algorithm is now experimentally tested for a MIMO case study using the same free-free beam excited by applying an input voltage to both MFC's simultaneously. The experimental setup and the procedure adopted follows a similar approach to the SIMO case with the only difference being multiple excitations. 

Uncorrelated input chirp voltage signals provided to MFC 1 and MFC 2 excite the beam simultaneously. The chirp is designed to have different cycles of frequency sweeping to remove the mode cancellation arising due to correlated input signals. The WDMD methodology builds a MIMO, data-driven, state-space model using the collected response measurements.  The learned model has state-space  dimensions ${\mathbf{A_w}} \in {\Re ^{140 \times 140}},{\mathbf{B_w}} \in {\Re ^{140 \times 2}},{\mathbf{C_w}} \in {\Re ^{10 \times 140}}$ and ${\mathbf{D_w}} \in {\Re ^{10 \times 2}}$ . The number of columns in matrix ${\mathbf{B}}$ matrix represents the number of inputs in the system, which in the present MIMO example is two. 
\begin{figure}[!htb]
\centering
\includegraphics[trim=0cm 0cm 0cm 0cm,scale=0.78]{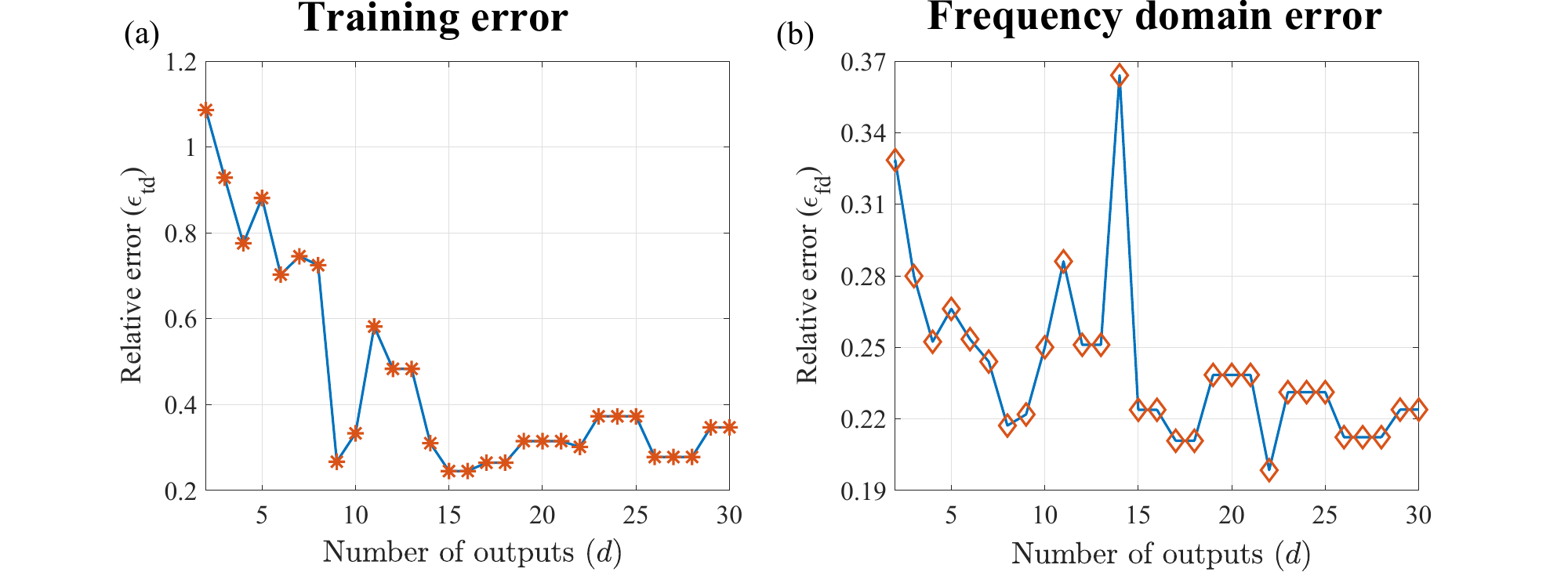}
\caption{Error convergence results for the experimental MIMO case study: (a) Relative time domain error for training phase and (b) Relative frequency domain error}
\label{mimo_conv}
\end{figure}

As in \Cref{simo}, the MIMO state-space model is excited with the training inputs to reproduce the results of the training phase.
For brevity, the time domain and frequency domain fitting results for the model are not shown.  \Cref{mimo_conv} shows the time domain error $\epsilon_{td}^{rel}$, and the frequency domain $\epsilon_{fd}^{rel}$ as a function of the number of outputs recorded and made available to the WDMD methodology. The lowest relative error value recorded is around $0.25$, which is slightly higher than the SIMO case. Similar is the case for $\epsilon_{fd}^{rel}$, which has a lowest recorded value of $0.19$, which is slightly higher than the SIMO case. 

\section{Conclusions and future work} \label{conc}

The current study presented a novel data-driven methodology to model dynamical systems from its input-output trajectories, without having access to governing physical equations or full internal state dynamics.
This was achieved using wavelets in conjunction with the ioDMD approach, leading to the proposed methodology, wavelet based DMD (WDMD). The numerical case study involving the dynamical response of a finite element cantilever beam was performed to demonstrate the effectiveness of WDMD. 
 WDMD  was utilized to develop a data-driven SIMO state-space  dynamical model of the FEM beam based on measured input-output response. 
{The WDMD methodology utilizes a subset of these measurements and approximates the underlying dynamics via a linear model using the maximal overlap discrete wavelet transform (MODWT) coefficients of the measured outputs as the auxiliary state-vector.} The error convergence studies illustrated that even with a few measured outputs, WDMD was able to model the underlying dynamical system accurately.
The experimental case study on a simple free-free beam, demonstrated the efficacy of WDMD methodology as an appropriate candidate for modeling practical dynamical systems despite having no access to internal state measurements.

The  work  presented  herein  demonstrates  the  feasibility  of  approximating the input-output dynamics of a vibrating beam based on measured input-output data using this new data-driven modeling approach. Although the WDMD algorithm performed reasonably in presence of noise, as demonstrated in \Cref{expstudty}, further studies into sensor placement, input excitation requirements and hyper-parameter selection are required to improve the robustness of the WDMD algorithm in the presence of higher noise level. 

\section*{ACKNOWLEDGEMENTS}\hspace*{\fill} \\
The authors would like to thank Drs.  Benner, Himpe,
and Mitchell  for providing access to their
input-output DMD code that has provided the foundation for building all the codes used for the current research. The authors acknowledge the support received through the Rolls Royce Fellowship and John R. Jones III Graduate Fellowship. Dr. Tarazaga acknowledges the support received through the John R. Jones III Faculty Fellowship. Gugercin was supported in parts by the National Science Foundation under Grant No. DMS-1819110.

\end{document}